\PassOptionsToPackage{table}{xcolor}
\documentclass[conference]{IEEEtran}
\let\labelindent\relax

\usepackage{amssymb}
\usepackage{amsmath}
\usepackage{tabularx}
\usepackage[breakable]{tcolorbox}
\usepackage{enumitem}

\usepackage{cite}

\usepackage{graphicx}
\usepackage{subcaption}
\usepackage{makecell}

\usepackage{multirow}

\newtheorem{definition}{Definition}
\newtheorem{theorem}{Theorem}

\begin{document}
%
\title{Boros: Secure Cross-Channel Transfers \\ via Channel Hub}


\author{\IEEEauthorblockN{YongJie Ye}
\IEEEauthorblockA{\textit{School of Data and Computer Science} \\
\textit{Sun Yat-sen University}\\
Guangzhou, China \\
yeyj6@mail2.sysu.edu.cn}
\and
\IEEEauthorblockN{Jingjing Zhang}
\IEEEauthorblockA{\textit{School of Data and Computer Science} \\
\textit{Sun Yat-sen University}\\
Guangzhou, China \\
zhangjj43@mail2.sysu.edu.cn}
\and
\IEEEauthorblockN{Weigang Wu}
\IEEEauthorblockA{\textit{School of Data and Computer Science} \\
\textit{Sun Yat-sen University}\\
Guangzhou, China \\
wuweig@mail.sysu.edu.cn}
\and
\IEEEauthorblockN{Xiapu Luo}
\IEEEauthorblockA{\textit{Department of Computing} \\
\textit{The Hong Kong Polytechnic University}\\
Hong Kong, China \\
csxluo@comp.polyu.edu.hk}
\and
\IEEEauthorblockN{Jiannong Cao}
\IEEEauthorblockA{\textit{Department of Computing} \\
\textit{The Hong Kong Polytechnic University}\\
Hong Kong, China \\
csjcao@comp.polyu.edu.hk}
}

\maketitle

\begin{abstract}
    The payment channel, which allows two parties to perform micropayments without involving the blockchain, 
    has become a promising proposal to improve the scalability of decentralized ledgers such as Bitcoin and Ethereum. 
    Payment channels have been extended to the payment network, through which users can utilize existing channels as intermediary links to route coins to others. 
    However, routing payments through multiple channels bears nontrivial overheads.
    It requires every intermediary channel to lock a portion of its available capacity until the payment is settled.
    This may lead to deadlock in a concurrent situation.
    The intermediary nodes in a payment path may also charge fees for routing a payment.
    The longer the routing path, the more serious the above problems.
    
    In this paper, we design and develop a novel off-chain system to shorten the routing path for the payment network. In particular,
    we propose the channel hub, which is an extension of the payment hub, to allows transferring coins directly from one payment channel to another within the same hub. That is, the channel hub can be viewed as a shortcut device for the underlying payment network.
    We design a new protocol named Boros to perform secure off-chain cross-channel transfers through the channel hub.
    We not only present the security definition of the Boros protocol formally but also prove its security using the UC-framework.
    To demonstrate the feasibility of the Boros protocol, 
    we develop a proof-of-concept prototype running on the Ethereum.
    Our evaluation shows that our system can effectively shorten the off-chain routing path.
\end{abstract}


%

\section{Introduction}

Since the advent of Bitcoin \cite{nakamoto2008bitcoin} in 2008, decentralized cryptocurrencies have gained great popularity over the last 10 years. 
The key innovation behind decentralized cryptocurrencies is the combination of consensus mechanisms and hash-linked chain of blocks. 
The use of consensus algorithms such as POW \cite{dwork1992pricing} and PBFT \cite{castro1999practical} makes it possible for all of its participants to maintain one single ledger without relying on any trusted third parties. 
The hash-linked chain of blocks, which is also called blockchain, boosts the computational requirements for adversaries trying to temper the block contents.
Each miner running the POW consensus algorithm is required to solve a computationally expensive hash puzzle.
The one who solves that puzzle was given the right to append a new block to the blockchain. 
Inspired by Bitcoin, other cryptocurrencies allowing users to develop and deploy their smart contracts, 
which are written in a Turing-complete programming language and support arbitrary complexity, have emerged.
The most prominent cryptocurrency that supports the execution of smart contracts is Ethereum \cite{wood2014ethereum}, 
which uses Solidity as the developing language of its smart contracts.

However, the deployment of globally consensus mechanism such as POW leads to serious scalability problem for decentralized cryptocurrencies.
In its current state, Bitcoin can only support up to 6$\sim$7 transactions per second while Ethereum supports up to 20 transactions per second \cite{croman2016scaling} and there is no order of magnitude growth of throughput with simple re-parameterization \cite{gervais2016security}. 
Such a low transaction throughput is far from enough to support the widespread use of decentralized cryptocurrencies. 
In contrast, Visa processes up to 47,000 transactions per second \cite{trillo2013stress}.

Recently many attempts have emerged to mitigate the scalability problem of decentralized cryptocurrencies such as 
alternative consensus mechanisms \cite{eyal2016bitcoin, luu2015scp, pass2017hybrid}, 
sharding \cite{zamani2018rapidchain, luu2016secure, kokoris-kogias2018omniledger}, 
usage of trusted execution environment \cite{lind2017teechain, zhang2016town}, 
sidechain \cite{back2014enabling, lerner2016drivechains}, 
and payment channels/networks \cite{decker2015fast, poon2016bitcoin, mccorry2016towards}, etc. 
In particular, payment channels allow two parties to perform micropayments privately without broadcasting all of them to the blockchain, thus improving the scalability of cryptocurrencies significantly.
To open a payment channel, two parties need to broadcast a funding transaction together with their deposits to the blockchain. 
After that, the payment channel is opened and the two parties can perform off-chain transactions securely without involving the blockchain.
The funding capacity of the opened payment channel is equal to the total deposits of these two parties. 
The off-chain transactions change the distribution of funds among the two parties.
At any point, each party can decide to close the payment channel by committing the final distribution of funds to the blockchain and get their cash back to their accounts.

Payment channels can be extended to payment networks\cite{poon2016bitcoin}.
Instead of conducting an on-chain transaction or establishing an expensive payment channel, 
one could utilize the so-called routed payment, which routes the payment over multiple existing intermediary payment channels, to transfer coins to others.
Efforts have been made to realize more efficient payment networks. 
For example, Sprites \cite{miller2017sprites} reduces the worst-case collateral time for off-chain payments. 
Others \cite{prihodko2016flare, malavolta2017concurrency, malavolta2017silentwhispers, roos2018settling, green2017bolt} focus on improving aspects like concurrency, security, and privacy for the routing process.

Different from routing packets in traditional data networks such as TCP/IP network, routing transactions in a payment network faces more challenges.
For example, routing a payment through multiple payment channels requires that each channel in the path has enough capacity for that payment.
Moreover, it also requires each intermediary channel to lock a portion of that channel's available collateral until the payment is settled.
This, however, may lead to deadlock in a concurrent situation \cite{malavolta2017concurrency}.
Another problem is that intermediary nodes in a payment path may charge fees for routing a payment.
Obviously, the longer the routing path, the more serious the above problems.



In this paper, we propose a novel off-chain system to shorten the payment path for the payment network.
First, 
we propose the notion of channel hub, which is an extension of the payment hub \cite{khalil2018nocust}.
The participants of channel hub vary from individual nodes to payment channels.
It allows coins to be directly transferred from one payment channel to another within the same channel hub.
Thus, the channel hub can be viewed as a shortcut device for the underlying payment network.
Compare with traditional node-level payment hub \cite{khalil2018nocust}, the channel hub does not require additional collaterals and allows deposits in the established payment channels to be reused.
Besides, it could also benefit more nodes at the same cost.

Second, based on the idea of channel hub, we design a new protocol named Boros to perform secure off-chain cross-channel transfers, which allows coins to be transferred between two parties.
The Boros protocol guarantees that an honest party will not bear any financial losses despite strong adversarial capabilities.
%
We not only present the security definition of Boros formally using the Universally Composable framework proposed by Canetti \cite{canetti2001universally} but also prove its security according to our definition.

Third, we develop a proof-of-concept prototype running on Ethereum to demonstrate the feasibility of the Boros protocol.
We measure the execution cost of each operation of the Boros protocol in payment networks of different sizes, and the experiemntal results show that our system can effectively reduce the average payment path length.

\subsubsection*{Organization of the paper}

In Section \ref{background}, we provide the necessary background and review the related studies on payment channels and payment networks.
We introduce the main idea of the Boros protocol in Section \ref{mainConstructionIdea} and present its formal security definition using the UC-framework is presented in Section \ref{formalDescription}
(Due to the page limit, formal security proof of the Boros protocol is provided in Appendix \ref{securityProof}).
%
Section \ref{implementation} reports our proof-of-concept implementation and the evaluation results of the Boros protocol on Ethereum. 
Finally, we conclude this paper in Section \ref{conclusion}.

\section{Background and Related Works} \label{background}

\subsection{Payment Channel}

Payment channels \cite{decker2015fast, poon2016bitcoin} allow parties to perform transfers privately without involving the blockchain, 
yet still keeping the ability for honest parties to reclaim its rightful amount of funds at any given time. 
Rather than committing each individual payment to the blockchain,
two parties broadcast a funding transaction together with their deposits to the blockchain to open a payment channel. 
The funding capacity of the opened payment channel is equal to the total deposits of these two parties. 
After the payment channel is successfully opened, these two parties can perform off-chain transfers securely without touching the blockchain.
The core of a payment channel protocol is to reach a consensus on the latest distribution of funds among the two parties and prevent malicious one from rolling back.
Decker et al. \cite{decker2015fast} use the blockchain based \textit{time locks} to invalidate obsolete distribution of funds 
while the Lightning Network \cite{poon2016bitcoin} relies on punishment to enforce the latest distribution.
At some point when any of them wishes to reclaim their funds, 
they broadcast a committing transaction, which contains the final distribution of funds, to the blockchain to close the payment channel and get their cash back to their account. 
Because all intermediary transfers are maintained only by these two parties and not required to be written to the blockchain, 
the payment channel can significantly increase the transaction throughput between two parties.
The network bandwidth is the only limitation of transaction rate. 

There are several improvement proposals on payment channel protocol.
Burchert et al. \cite{burchert2018scalable} introduce a new layer called channel factory between the blockchain and the payment network so that it can quickly refund a payment channel.
Green et al. \cite{green2017bolt} propose Bolt for constructing privacy-preserving payment channels while lowering the storage burden on the payment network.
Dziembowski et al. \cite{dziembowski2017perun} design Perun to establish a virtual payment channel between two parties that are connected by one intermediary.
The virtual payment channel allows these two parties to perform transfers and do not require the intermediary to confirm every individual payment.
This can significantly reduce latency and costs while improving privacy since the intermediary cannot observe the individual transfers between the two parties.
The state channel \cite{dziembowski2018general} allows off-chain execution of arbitrary complex smart contracts.
They also propose a novel technique to recursively build virtual state channel that spans multiple ledgers or virtual state channels.

\subsection{Payment Network}

Instead of opening an expensive payment channel or conducting on-chain transactions, 
two parties without direct connection by a payment channel can utilize existing channels as intermediary links to route coins over the payment network \cite{poon2016bitcoin, mccorry2016towards}.

The most critical challenge when routing a payment through multiple intermediary channels is to enforce atomicity.
That is, either the capacity of all intermediary channels in the path is updated or none of them is changed.
To securely conduct transfers across multiple payment channels, the Lightning Network \cite{poon2016bitcoin} adopts a technique called \textit{Hash Time-Lock Contract} (HTLC).
An HTLC is a conditional contract where the condition is enforced by the blockchain so it does not require trust in any participant in the network.
This contract locks a portion of coins that can be released by its receiver only if the condition is fulfilled or returned to its owner if the contract times out.
When routing a payment over the payment network, the receiver generates a secret value R and sends the hash value of R, denotes as $y$ where $y=H(R)$, to all intermediary nodes in the path.
Each intermediary channel then sets up an HTLC using the hash value $y$ to lock a portion of its coins, which is equal to the payment amount plus some optional routing fees.
Finally, the receiver discloses the secret value R to finish that payment and release the locked coins at each intermediary channel.


Recent studies have further improved the payment networks.
Sprites \cite{miller2017sprites} reduces the worst-case collateral time for off-chain payments. 
Malavolta et al. \cite{malavolta2017concurrency} proposes the first non-blocking protocol for the payment network, where at least one out of a set of concurrent payments can finally complete, and gives an in-depth discussion on the trade-off between privacy and concurrency.
Revive \cite{khalil2017revive} is the first rebalance scheme for the off-chain payment network, which enables a set of members in a skewed payment channel network to safely shift balances between their payment channels to reach a balanced state. 
There are also several works that focus on improving efficiency and privacy of the routing process in the decentralized payment network such as Flare \cite{prihodko2016flare}, SilentWhispers \cite{malavolta2017silentwhispers}, and SpeedyMurmurs \cite{roos2018settling}.

\subsection{Payment Hub}

TumbleBit \cite{ethan2017tumblebit} introduces the concept of payment hub which allows a payer to perform secure off-chain payment to a set of payees within the same payment hub. 
Payments are performed off-chained with the help of an untrusted intermediary called the Tumbler.
It guarantees that no one, not even the Tumbler can violate anonymity and link a payment from its payer to its payee.
TumbleBit is fully compatible with Bitcoin protocol.
However, TumbleBit requires that the Tumbler opens a directed payment channel with each participant.
This, however, would lead to fragmented collaterals and significantly complicate the operation of the payment hub.

Khalil et al. \cite{khalil2018nocust} propose NOCUST, which allows the collaterals of its participants to be centrally managed in bulk and thus significantly reduces the operating costs of the payment hub.
A NOCUST payment hub consists of two fundamental components: an on-chain verifier contract and an off-chain operator server.
The on-chain verifier contract serves as a trusted financial custodian.
It maintains collaterals of all participants of the payment hub and is responsible for resolving disputes.
The off-chain operator server executes every transfer and synchronizes with the on-chain verifier contract periodically to keep consistency.
NOCUST guarantees that an honest participant can always maintain custody of its funds and its enacted transfers can be finally delivered.


However, the node-level payment hub does not allow reusing the deposits in the established payment channels.
If a node wishes to join in a payment hub, it has to come up with additional collaterals instead of reusing existing ones in its payment channels.
In this work, we extend the concept of the payment hub to channel hub, whose participants vary from individual nodes to payment channels.
It allows transferring coins from one payment channel to another within the same channel hub.
After joining a channel hub, the deposit of a payment channel can be used for both cross-channel transfers and traditional in-channel transfers. It is worth noting that compared with the payment hub, the channel hub can benefit more nodes in the same cost. More precisely, both the payment hub and the channel hub require one on-chain transaction for a participant to join in.
However, since the participant of the channel hub is payment channel, it allows both parties of the payment channel to benefit from the channel hub.

\section{Main Construction Idea} \label{mainConstructionIdea}

\subsection{Channel Hub} \label{channelHub}

We adopt the construction of NOCUST \cite{khalil2018nocust} and extend the concept of payment hub to channel hub, which allows transferring coins from one payment channel to another within the same channel hub.
Here, we only give a general description of the payment hub and describe the differences between the payment hub and our channel hub. For further details, we refer the reader to \cite{khalil2018nocust}.

A payment hub is composed of two basic components: 
an on-chain verifier smart contract $\mathcal{V}_{\not\subset}$ and an off-chain operator server $\mathcal{O}_{\not\subset}$. 
The off-chain operator server is an interactive server that acts as a financial intermediary, i.e., all off-chain transfers performed within the payment hub need to be relayed and ratified by the operator server. 
Meanwhile, the operator server maintains a local ledger $\mathcal{B}^{\mathcal{L}}$ which contains the balance of its participants and all information related to the transfers performed through $\mathcal{O}_{\not\subset}$. 
The information in $\mathcal{B}^{\mathcal{L}}$ is periodically committed to the global ledger $\mathcal{B}^{\mathcal{G}}$, which is maintained by the on-chain verifier $\mathcal{V}_{\not\subset}$,  to keep global consistency.
Apart from maintaining the global ledger $\mathcal{B}^{\mathcal{G}}$, the on-chain verifier $\mathcal{V}_{\not\subset}$ also serves as a dispute resolver in case of malicious participants or even dishonest operator server to guarantee the balance custody of honest participants and enforce enacted transfers.

The core of the payment hub is a mapping: $\{0, 1\}^{\lambda} \to \mathbb{N}_0$, where the fixed-length binary string $\{0, 1\}^{\lambda}$ denotes the account of each participant and $\mathbb{N}_0$ denotes its balance.
The key observation motivating our extension is that the accounts of both individual nodes (known as external account) and payment channels (known as contract account) share the same address space.
In other words, both of the external account and the contract account are represented by a fixed-length binary string $\{0, 1\}^{\lambda}$.
Thus, our extension can be simply accomplished by letting $\{0, 1\}^{\lambda}$ denote the contract account of the payment channel and $\mathbb{N}_0$ denote its funding capacity.
Concretely, we modify the \textit{Merklelized interval tree} $\mathcal{T}_{\not\subset}$ data structure used in NOCUST \cite{khalil2018nocust}
so that 
each leaf of the tree $\mathcal{T}_{\not\subset}$ stores the information corresponds to a payment channel $\beta_i$, 
which mainly consists of the channel contract account $\alpha_i$, the funding capacity $c_i$, and the last update $u_i$ that channel $\beta_i$ involved in.

\subsection{Informal Description of Boros Protocol} \label{informalDescription}

In this section, we informally describe the basic idea of Boros protocol, which uses the channel hub to perform secure point-to-point cross-channel transfers.
The Boros protocol is designed to prevent any honest node from losing funds despite strong set of adversarial capabilities. 

Suppose A wishes to transfer $\Delta x$ coins to B. 
This can be done by first transferring $\Delta x$ coins from channel $\beta_{AC}$ to channel $\beta_{BD}$ using the channel hub, 
and then updating the distribution of deposits in channel $\beta_{AC}$ and $\beta_{BD}$, so that $\beta_{AC}.balance(A) \mathrel{{-}{=}} \Delta x$ and $\beta_{BD}.balance(B) \mathrel{{+}{=}} \Delta x$, while the balances of C and D remain unchanged. 
The key issue is to enforce balance consistency in the whole process. 

\begin{figure}[ht]
  \centering
    \includegraphics[width=0.48\textwidth] {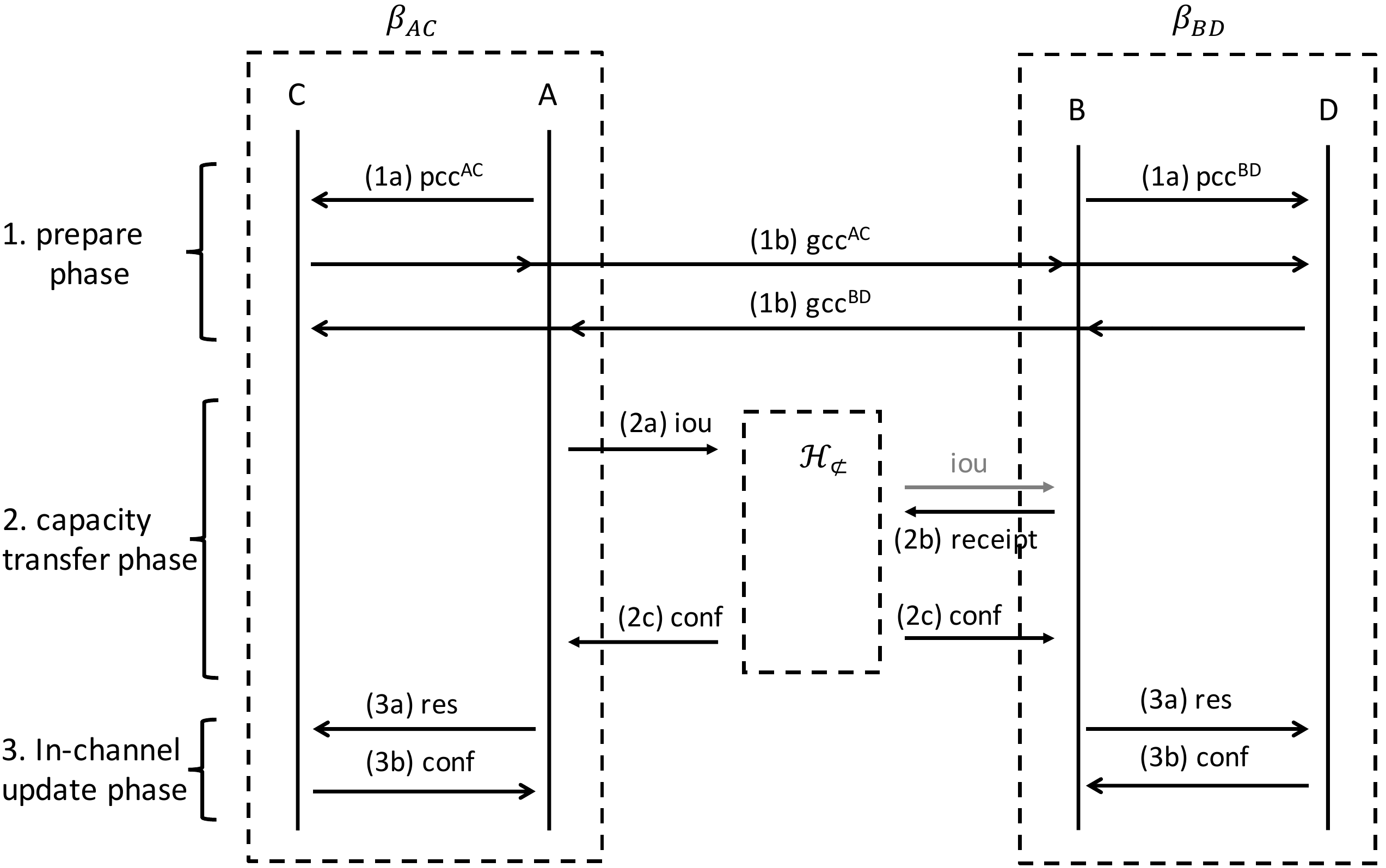}
    \caption {Outline for the cross-channel transfer with Boros.}
    \label{fig:mci:3}
\end{figure}

We first describe the whole process of cross-channel transfer in case all parties are honest. 
Figure \ref{fig:mci:3} shows the messages flow of the protocol.
We assume that both channel $\beta_{AC}$ and channel $\beta_{BD}$ have already joined in the same channel hub $\mathcal{H}_{\not\subset}$. 
The Boros protocol consists of three phases, namely the prepare, capacity transfer, and in-channel update phase. 

In the prepare phase, A sends a message $m_{\text{pcc}}$ to C, indicating that A wishes to transfer his $\Delta x$ coins from channel $\beta_{AC}$ to channel $\beta_{BD}$ while keeping C's balance on $\beta_{AC}$ unchanged. 
The word ``pcc'' is the abbreviation for ``Prepare Cross-Channel transfer''.
When C receives $m_{\text{pcc}}$, C will check the validity of message $m_{\text{pcc}}$ and broadcast a message $m_{\text{gcc}}$ to both A, B, and D, indicating that C agrees with that cross-channel transfer. 
The word ``gcc'' is the abbreviation for ``Grant Cross-Channel transfer''. 
The interactions between B and D are handled analogously.
At the end of the prepare phase, A, C, B and D should hold the messages $m_{\text{gcc}}^{AC}$ and $m_{\text{gcc}}^{BD}$.

In the capacity transfer phase, A and B, on behalf of channel $\beta_{AC}$ and channel $\beta_{BD}$ respectively, perform the coin transfer between channel $\beta_{AC}$ and channel $\beta_{BD}$ via the channel hub $\mathcal{H}_{\not\subset}$.
Here we follow the operations of NOCUST \cite{khalil2018nocust}.
First, A sends $m_{\text{iou}}$ to the off-chain operater server $\mathcal{O}_{\not\subset}$. $\mathcal{O}_{\not\subset}$ then checks the validity of $m_{\text{iou}}$, notifies B and waits for B's receipt. 
When B is notified by $\mathcal{O}_{\not\subset}$, it will also verify the validity of $m_{\text{iou}}$ and then reply with a signed receipt. 
Upon receiving B's receipt, $\mathcal{O}_{\not\subset}$ confirms the IOU execution and sends $m_{\text{conf}}$ to both A and B. 
At that point, the capacity transfer phase is completed. 
The funding capacity of channel $\beta_{AC}$ decreases by $\Delta x$ and $\beta_{BD}$ increases by $\Delta x$.

In the in-channel update stage, both channel $\beta_{AC}$ and channel $\beta_{BD}$ need to update the distribution of balance to keep consistency. 
Concretely, in channel $\beta_{AC}$, A's balance should be decreased by $\Delta x$ while C's balance remains unchanged; in channel $\beta_{BD}$, B's balance should be increased by $\Delta x$ while D's balance remains unchanged. 
Taking channel $\beta_{AC}$ as an example, in this phase, A sends C a message $m_{\text{icu}}$ indicating the result of the capacity transfer phase to C. 
The word ``icu'' stands for ``In-Channel Update''.
When the capacity transfer phase successes, the message $m_{\text{icu}}$ will contain information about the decrease on A's balance. Otherwise, $m_{\text{icu}}$ will tell C not to change A's balance.
Upon receiving the $m_{\text{icu}}$, C checks the validity of that message then reply with a confirmation $m_{\text{conf}}$. 
The interactions between B and D are handled in a similar way and then the whole transfer is completed.

\subsection{Security Properties} \label{securityProperties}

In this section, we describe the threat model and the security properties of our protocol. 

\subsubsection*{Threat model}
We assume the presence of irrational adversary willing to lose some or even all of its funds to cause honest parties to bear financial losses.
This irrational adversary may take control of some or even all but one of the participants involved in a cross-channel transfer. The internal state and all of the following communications of the corrupted party are  exposed to the adversary. Besides, the adversary may send arbitrary messages on behalf of the corrupted party.
On the other hand, we assume that the communication channels between honest parties and the integrity of the honest parties' identity can not be corrupted by the adversary. \\

Against the above threat model, our protocol guarantees the following security properties:

\begin{itemize}[leftmargin=*]
    \item \textbf{Consensus on channel hub enrollment and withdrawal.} Our protocol guarantees that honest parties can always reach a consensus on whether a payment channel has joined in or withdrawn from a channel hub.
    \item \textbf{Consensus on channel capacity.} Our protocol guarantees that honest parties can always learn the funding capacity of their payment channels. That is, an honest party can always learn the result of every cross-channel transfer involving it.
    \item \textbf{Balance security.} Intuitively, balance security guarantees that an honest party will not lose any of his coins despite strong adversarial capabilities, i.e., an honest party will not bear financial losses even when all other participants involved in a cross-channel transfer are malicious.
\end{itemize}

\subsection{Concise Proof of Misbehavior}


We now investigate what happens if some of the parties are malicious. 
Let's first consider the prepare phase, whose main purpose is to reach an agreement about the following transfer among A, C, B, and D. 
That is, all of these four parties should receive both $m_{\text{gcc}}^{AC}$ and $m_{\text{gcc}}^{BD}$ at the end of this phase.
If there exist malicious parties that do not send or reply messages, then there must be someone failed to collect both of these two messages.
Now we discuss the following cases: 
(1) A cannot obtain both of these messages. 
(2) B fails to collect both of these messages. 
(3) C or D cannot collect both of these messages.

For case (1), the consequences are obvious. 
If A cannot collect both $m_{\text{gcc}}^{AC}$ and $m_{\text{gcc}}^{BD}$, then he will fail to construct the message $m_{\text{iou}}$ in next phase, which leading to the termination of the whole transfer. 
For case (2), B is quite tolerant since B can still get $m_{\text{gcc}}^{AC}$ and $m_{\text{gcc}}^{BD}$ from the $m_{\text{iou}}$ when he got notified by the channel hub $\mathcal{H}_{\not\subset}$.

The case (3) is much more complicated. 
In this case, C or D cannot determine whether this transfer is prepared to be performed.
In the worst case, they fail to learn whether the transfer is performed or not. 
In order to eliminate the feasibility of being inconsistent, we do not allow a party to involve in multiple transfers at the same time. 
Besides, we introduce $\mathcal{T}$ to denote the maximum transfer duration. 
When a transfer starts, a party should learn the result before $\mathcal{T}$. 
Otherwise, it will complain to the channel hub.
For example, if C cannot get the result of that transfer from A, then he waits until $\mathcal{T}$ expires and sends a force-reply message $m_{\text{fr}}$ to $\mathcal{H}_{\not\subset}$.
Once receiving message $m_{\text{fr}}$, $\mathcal{H}_{\not\subset}$ informs A about C's complaint, asking A to provide the result $m_{\text{icu}}$ within a fixed time $\Delta$. 
If A replies with a valid $m_{\text{icu}}$ containing right amount of funding capacity of channel $\beta_{AC}$ in time, then $\mathcal{H}_{\not\subset}$ forwards $m_{\text{icu}}$ to C such that C can proceed. 
Otherwise, $\mathcal{H}_{\not\subset}$ considers channel $\beta_{AC}$ as unresponsive and then closes it, refunding their cash according to the result of that cross-channel transfer. 

In the capacity transfer phase, coins are transferred from channel $\beta_{AC}$ to channel $\beta_{BD}$ via the channel hub $\mathcal{H}_{\not\subset}$. 
NOCUST \cite{khalil2018nocust} guarantees that an honest party P can always maintain custody of its funds and ensure that its enacted transfers are correctly delivered within the hub. 
Such guarantees hold for our channel hub under the same attacker model.
In a nutshell, NOCUST allows a party P to open the so-called ``balance update challenge'' and ``transfer delivery challenge'' to the on-chain verifier contract $\mathcal{V}_{\not\subset}$ to enforce secure guarantees. 
The former challenges against the integrity of P's balance and the latter challenges against the integrity of an off-chain transfer deliver in the hub.
For further details on security analysis, we refer the reader to \cite{khalil2018nocust}.

In the in-channel update phase, both channel $\beta_{AC}$ and channel $\beta_{BD}$ need to update its distribution of balance to keep consistency. 
Possible exceptions that may occur in this phase include: 
(1) A is dishonest and does not send the result $m_{\text{icu}}$ to C; and  
(2) C is malicious and does not reply A with confirmation $m_{\text{conf}}$.
For case (1),  A is dishonest and deliberately conceals the result of the capacity transfer phase from C, and hence C is unable to determine the funding capacity of channel $\beta_{AC}$.
As discussed in the prepare phase, in this situation, C simply waits until $\mathcal{T}$ expires and then complain to $\mathcal{H}_{\not\subset}$.
For case (2), C gets $m_{\text{icu}}$ from A but doesn't reply with confirmation $m_{\text{conf}}$. 
This situation is similar to (1). To resolve this issue, A just waits until $\mathcal{T}$ expires and then contacts $\mathcal{H}_{\not\subset}$ to enforce C's confirmation.

\section{Formal Description} \label{formalDescription}

\subsubsection*{Universally Composable model}

In the UC model \cite{canetti2001universally}, the security of a protocol is defined by comparing the execution of the protocol in the real-world model with an ideal process.
In real-world model, the n-party protocol $\pi$ is executed by a set of parties $P \in \{P_1, P_2, ..., P_n\}$, which is modeled as probabilistic polynomial time (PPT) machine. 
In real-world there exists an adversary $\mathcal{A}$, who can corrupt some of these parties such that the internal state and all the future actions of the corrupted party are totally controlled by the adversary. 
For simplicity, we only consider static corruption, which means that the adversary $\mathcal{A}$ must decide which parties to corrupt at the beginning of protocol execution. 
Both the parties and the adversary $\mathcal{A}$ receive inputs and output to the environment $\mathcal{Z}$, which is used to model all factors that are external to the current protocol execution.
In the ideal-world, the environment $\mathcal{Z}$ interacts with the ideal functionality $\mathcal{F}$ via the so-called dummy parties, who simply forward messages from $\mathcal{Z}$ to $\mathcal{F}$ and back.
The counterpart of the adversary $\mathcal{A}$ in the ideal-world is the simulator $\mathcal{S}$.
Then we say a protocol $\pi$ is considered secure if the environment $\mathcal{Z}$ can not distinguish whether it is interacting with $\mathcal{A}$ and $\pi$ running in real-world model or with $\mathcal{S}$ and ideal functionality $\mathcal{F}$ in ideal process.

\subsubsection*{Communication model}

For the sake of simplicity, we assume a synchronous communication model in which all parties proceeds in synchronous round and all parties start simultaneously. 
In this model, every party can send messages to all other parties and the message sent in round $i$ arrives at its destination at the beginning of round $i+1$.
When it comes to the ideal functionalities, we simply assume that the computation of ideal functionalities and communication with ideal functionalities are instantaneous.
The synchronous communication model can be achieved by a global clock functionality. For further details, we refer to \cite{katz2013universally, kiayias2016fair, hofheinz2004synchronous}.

\subsubsection*{The ledger functionality $\mathcal{F}_{\mathcal{L}}$}

Following \cite{dziembowski2018fairswap}, we model the global ledger as an ideal functionality $\mathcal{F}_{\mathcal{L}}$. 
The internal state of $\mathcal{F}_{\mathcal{L}}$ consists of a public-accessed account space denoted as $\mathcal{B} : \alpha_i \to p_i$, 
where $\alpha_i \in \{0, 1\}^{\lambda}$ denotes either an external account or contract account, and $p_i \in \mathbb{N}_0$ denotes the balance of account $\alpha_i$.
The ledger functionality $\mathcal{F}_{\mathcal{L}}$ provides the following interface:
\begin{itemize}[leftmargin=*]
    \item \textit{transfer}, which allows to transfers $p$ coins from account $\alpha_i$ to $\alpha_j$ via sending message $(\text{transfer}, sid, \alpha_i, \alpha_j, p)$.
\end{itemize}
To simplify notation, we assume that every ideal functionality $\mathcal{F}_{f}$ has a special account $\alpha_{f}$.
When we say that the ideal functionality $\mathcal{F}_{f}$ receives a message $m$ together with $p$ coins from A, we actually mean that upon receiving message $m$, the ideal functionality $\mathcal{F}_{f}$ sends a message $(\text{transfer}, sid, \alpha_A, \alpha_f, p)$ to $\mathcal{F}_{\mathcal{L}}$. 
Similarly, when we say that the ideal functionality $\mathcal{F}_{f}$ sends $p$ coins back to A, we actually mean that $\mathcal{F}_{f}$ sends a message $(\text{transfer}, sid, \alpha_f, \alpha_A, p)$ to $\mathcal{F}_{\mathcal{L}}$. 
The \textit{transfer} interface also allows the simulator $\mathcal{S}$ to simulate the ``irrational'' parties who are willing to sacrifice their funds to cause honest parties to lose some or all of their funds.
In such cases, we simply let $\mathcal{S}$ to transfer coins from the account of the irrational party to the honest one.
We note that the ideal functionality $\mathcal{F}_{\mathcal{L}}$ mentioned above only captures the basic ideal concept of the global ledger for the convenience of exposition. 
A more accurate and realistic formalization of the global ledger can be found in \cite{kosba2016hawk}.

\subsubsection*{The channel hub functionality $\mathcal{F}_{\mathcal{N}}$}

\begin{figure}[ht]
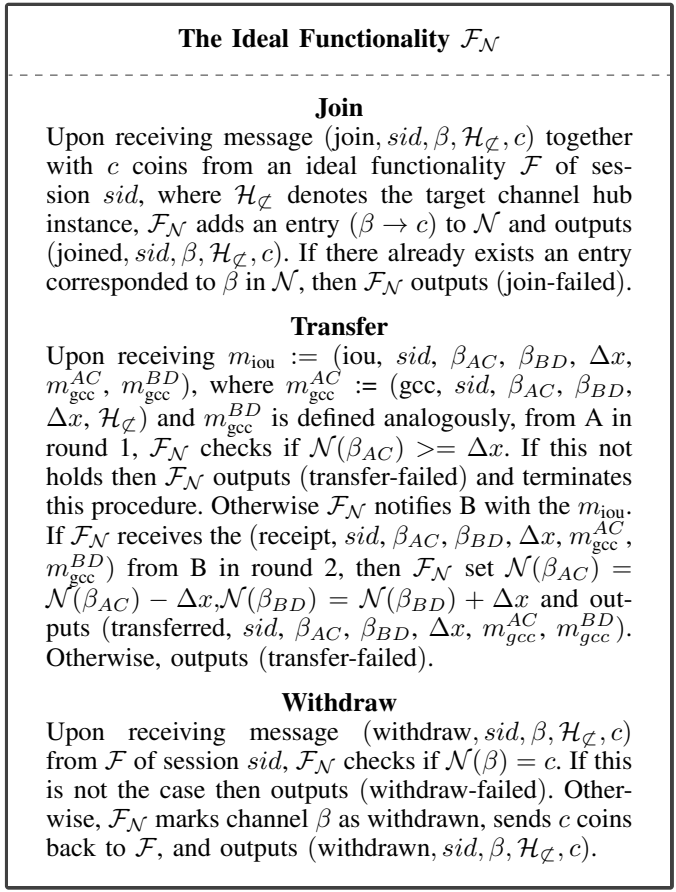

    \begin{tcolorbox}[colback=white, arc=0pt]
    
        \centerline {\textbf{The Ideal Functionality $\mathcal{F}_{\mathcal{N}}$}}
    
        \tcblower
    
        \centerline {\textbf{Join}}
        
        Upon receiving message $(\text{join}, sid, \beta, \mathcal{H}_{\not\subset}, c)$ together with $c$ coins from an ideal functionality $\mathcal{F}$ of session $sid$, 
        where $\mathcal{H}_{\not\subset}$ denotes the target channel hub instance, 
        $\mathcal{F}_{\mathcal{N}}$ adds an entry $(\beta \to c)$ to $\mathcal{N}$ and outputs $(\text{joined}, sid, \beta, \mathcal{H}_{\not\subset}, c)$.
        If there already exists an entry corresponded to $\beta$ in $\mathcal{N}$, 
        then $\mathcal{F}_{\mathcal{N}}$ outputs $(\text{join-failed})$.
    
        \medskip
        \centerline {\textbf{Transfer}}
    
        Upon receiving $m_{\text{iou}} := (\text{iou}$, $sid$, $\beta_{AC}$, $\beta_{BD}$, $\Delta x$, $m_{\text{gcc}}^{AC}$, $m_{\text{gcc}}^{BD})$, where $m_{\text{gcc}}^{AC}$ := $(\text{gcc}$, $sid$, $\beta_{AC}$, $\beta_{BD}$, $\Delta x$, $\mathcal{H}_{\not\subset})$ and $m_{\text{gcc}}^{BD}$ is defined analogously, from A in round 1, $\mathcal{F}_{\mathcal{N}}$ checks if $\mathcal{N}(\beta_{AC}) >= \Delta x$. If this not holds then $\mathcal{F}_{\mathcal{N}}$ outputs $(\text{transfer-failed})$ and terminates this procedure. Otherwise $\mathcal{F}_{\mathcal{N}}$ notifies B with the $m_{\text{iou}}$. \\
        If $\mathcal{F}_{\mathcal{N}}$ receives the $(\text{receipt}$, $sid$, $\beta_{AC}$, $\beta_{BD}$, $\Delta x$, $m_{\text{gcc}}^{AC}$, $m_{\text{gcc}}^{BD})$ from B in round 2, then $\mathcal{F}_{\mathcal{N}}$ set $\mathcal{N}(\beta_{AC}) = \mathcal{N}(\beta_{AC}) - \Delta x$$, $$\mathcal{N}(\beta_{BD}) = \mathcal{N}(\beta_{BD}) + \Delta x$ and outputs $(\text{transferred}$, $sid$, $\beta_{AC}$, $\beta_{BD}$, $\Delta x$, $m_{gcc}^{AC}$, $m_{gcc}^{BD})$. Otherwise, outputs $(\text{transfer-failed})$.
    
        \medskip
        \centerline {\textbf{Withdraw}}
        
        Upon receiving message $(\text{withdraw}, sid, \beta, \mathcal{H}_{\not\subset}, c)$ from $\mathcal{F}$ of session $sid$, $\mathcal{F}_{\mathcal{N}}$ checks if $\mathcal{N}(\beta) = c$. 
        If this is not the case then outputs $(\text{withdraw-failed})$. 
        Otherwise, $\mathcal{F}_{\mathcal{N}}$ marks channel $\beta$ as withdrawn, sends $c$ coins back to $\mathcal{F}$, and outputs $(\text{withdrawn}, sid, \beta, \mathcal{H}_{\not\subset}, c)$.

    \end{tcolorbox}
    \caption{The ideal functionality $\mathcal{F}_{\mathcal{N}}$ for channel hub}
    \label{FN}
\end{figure}

In Figure \ref{FN}, we outline the ideal functionality of the channel hub $\mathcal{F}_{\mathcal{N}}$. 
The channel hub functionality $\mathcal{F}_{\mathcal{N}}$ maintains a channel space denoted as $\mathcal{N}: \beta_i \to c_i$, 
where $\beta_i \in \{0, 1\}^\lambda$ denotes the contract account of channel $\beta_i$, and $c_i \in \mathbb{N}_0$ denotes the funding capacity of channel $\beta_i$.
When we say that a payment channel $\beta$ is marked as joined, we mean that an entry corresponded to channel $\beta$ is added to $\mathcal{N}$.
Similarly, when we say that a payment channel $\beta$ is marked as withdrawn, we mean that the entry corresponded to channel $\beta$ is removed from $\mathcal{N}$.
The channel hub functionality $\mathcal{F}_{\mathcal{N}}$ provides the \textit{join} and \textit{withdraw} interfaces for other ideal functionalities and the interface \textit{transfer} for parties.
(1) \textit{join}: When triggered by a joining request together with $c$ coins from an ideal functionalities, where $c$ denotes the funding capacity of the channel $\beta$, the ideal functionality $\mathcal{F}_{\mathcal{N}}$ marks the channel $\beta$ as joined. 
(2) \textit{transfer}: Upon receiving an iou request $(\text{iou}, \beta_{AC}, \beta_{BD}, \Delta x)$ from A and a signed receipt from B, $\mathcal{F}_{\mathcal{N}}$ moves $\Delta x$ coins from channel $\beta_{AC}$ to $\beta_{BD}$. 
(3) \textit{withdraw}: a payment channel $\beta_i$ can withdraw from a channel hub at any time by sending a withdrawal request to $\mathcal{F}_{\mathcal{N}}$.
The ideal functionality $\mathcal{F}_{\mathcal{N}}$ will eventually send $c_i$ coins back to the contract account of $\beta$ and marks channel $\beta_i$ as withdrawn.

\subsubsection*{The contract functionality $\mathcal{F}_{\mathcal{C}}$}

In Figure \ref{FC}, we outline the contract functionality $\mathcal{F}_{\mathcal{C}}$, which is the ideal functionality of the payment channel contract deployed on the blockchain.
The contract functionality $\mathcal{F}_{\mathcal{C}}$ maintains the set of active contract instance.
Each contract instance corresponds to a payment channel.
A contract instance is created when a payment channel is opened and removed when the payment channel is closed.
The contract functionality $\mathcal{F}_{\mathcal{C}}$ provides \textit{open}, \textit{join}, \textit{withdraw}, and \textit{close} interfaces for parties.

\subsubsection*{UC definition of security}

Let Boros be a protocol with access to the global ledger functionality $\mathcal{F}_{\mathcal{L}}$, 
the channel hub ideal functionality $\mathcal{F}_{\mathcal{N}}$, 
and the contract functionality $\mathcal{F}_{\mathcal{C}}$. 
The output of an environment $\mathcal{Z}$ interacting with Boros and an adversary $\mathcal{A}$ on security parameter $\lambda \in \mathbb{N}$, and auxiliary input $x \in \{0, 1\}^\ast$ is denoted as 
$EXEC^{\mathcal{F}_{\mathcal{L}}, \mathcal{F}_{\mathcal{N}}, \mathcal{F}_{\mathcal{C}}}_{\text{Boros}, \mathcal{A}, \mathcal{Z}}(\lambda, x)$.
In ideal world, we use 
$IDEAL^{\mathcal{F}_{\mathcal{L}}, \mathcal{F}_{\mathcal{N}}, \mathcal{F}_{\mathcal{C}}}_{\mathcal{F}_{\mathcal{H}}, \mathcal{S}, \mathcal{Z}}(\lambda, x)$
to denote the output of $\mathcal{Z}$ runing with the ideal functionality $\mathcal{F}_{\mathcal{H}}$ and the simulator $\mathcal{S}$.

\begin{definition}
Let $\lambda \in \mathbb{N}$ be a security parameter and $x \in \{0, 1\}^\ast$ be an auxiliary input, Boros be a protocol runing in the ($\mathcal{F}_{\mathcal{L}}$, $\mathcal{F}_{\mathcal{N}}$, $\mathcal{F}_{\mathcal{C}}$)-hybrid world. We say that protocol Boros realize the ideal functionality $\mathcal{F}_{\mathcal{H}}$ if for every adversary $\mathcal{A}$ there exists a simulator $\mathcal{S}$ such that for all PPT environments $\mathcal{Z}$:
\[
    EXEC^{\mathcal{F}_{\mathcal{L}}, \mathcal{F}_{\mathcal{N}}, \mathcal{F}_{\mathcal{C}}}_{\text{Boros}, \mathcal{A}, \mathcal{Z}}(\lambda, x) \approx IDEAL^{\mathcal{F}_{\mathcal{L}}, \mathcal{F}_{\mathcal{N}}, \mathcal{F}_{\mathcal{C}}}_{\mathcal{F}_{\mathcal{H}}, \mathcal{S}, \mathcal{Z}}(\lambda, x)
\]
where ``$\approx$'' denotes computational indistinguishability.

\end{definition}

\subsection{Ideal Functionality $\mathcal{F}_{\mathcal{H}}$ for Boros}

\begin{figure*}
    \begin{tcolorbox}[colback=white, arc=0pt]
    
        \centerline {\textbf{Ideal functionality $\mathcal{F}_{\mathcal{H}}$}}
    
        \tcblower
    
        \centerline{\textbf{Open a payment channel}}
        Upon receiving $(\text{open}, sid, \beta_{AC}, x_a)$ together with $x_a$ coins from A in round 1 and $(\text{open}, sid, \beta_{AC}, x_c)$ together with $x_c$ coins from C in round 2, $\mathcal{F}_{\mathcal{H}}$ adds a new entry $(\beta_{AC} \to \{c, \theta_1\})$ to $\mathcal{B}$, where $c = x_a + x_c$, $\theta_1(A) = x_a$, and $\theta_1(C) = x_c$, and outputs $(\text{opened}, sid, \beta_{AC}, c, \theta_1)$. Otherwise, $\mathcal{F}_{\mathcal{H}}$ sends $x_a$ coins back to A and outputs $(\text{open-failed})$.
    
        \medskip
        \centerline{\textbf{In-channel transfer}}
        Upon receiving $(\text{update}, sid, \beta_{AC}, \theta_{w+1})$ from A in round 1, where $\theta_{w+1}$ denotes a new distribution function that reflects the result of the update, $\mathcal{F}_{\mathcal{H}}$ sends a message $(\text{updating}, sid, \beta_{AC}, \theta_{w+1})$ to C.
        If next round C replies with $(\text{update-ok}, sid, \beta_{AC}, \theta_{w+1})$, then $\mathcal{F}_{\mathcal{H}}$ sets $(\beta_{AC} \to \{c, \theta_{w+1}\})$ and outputs $(\text{updated})$.
    
        \medskip
        \centerline{\textbf{Join a channel hub}}
        Upon receiving $(\text{join}, sid, \beta_{AC}, \mathcal{H}_{\not\subset})$ from A in round 1, where $\mathcal{H}_{\not\subset}$ denotes the target channel hub to join, $\mathcal{F}_{\mathcal{H}}$ checks if $\beta_{AC} \in \mathcal{B}_{\not\subset}$. 
        If this is the case then $\mathcal{F}_{\mathcal{H}}$ ignores the joining request and terminates this procedure. 
        Otherwise $\mathcal{F}_{\mathcal{H}}$ sends a message $(\text{join}, sid, \beta_{AC}, \mathcal{H}_{\not\subset}, c, \theta_w, \sigma_{A})$ where $\sigma_{A}$ denotes A's signature on that message to C. 
        If next round C replies with $(\text{join}, sid, \beta_{AC}, \mathcal{H}_{\not\subset}, c, \theta_w, \sigma_{A}, \sigma_{C})$, then $\mathcal{F}_{\mathcal{H}}$ marks channel $\beta_{AC}$ as joined and outputs $(\text{joined})$. Otherwise, $\mathcal{F}_{\mathcal{H}}$ outputs $(\text{join-failed})$.
    
        \medskip
        \centerline{\textbf{Cross-channel transfer}}
    
        \begin{itemize}[leftmargin=*]
            \item{\textbf{Prepare phase}} \\
            Upon receiving $(\text{cc-transfer}, sid, \beta_{AC}, \beta_{BD}, \Delta x, \mathcal{H}_{\not\subset})$ from A in round 1, 
            $\mathcal{F}_{\mathcal{H}}$ checks if the following three conditions hold: 
            (1) $\beta_{AC} \in \mathcal{B}_{\not\subset}$, 
            (2) $\beta_{BD} \in \mathcal{B}_{\not\subset}$, and 
            (3) $\theta_w(A) \geq \Delta x$. 
            If any of these conditions is not met then $\mathcal{F}_{\mathcal{H}}$ ignores the message and terminates this procedure. 
            Otherwise, $\mathcal{F}_{\mathcal{H}}$ sends $(\text{pcc}, sid, \beta_{AC}, \beta_{BD}, \Delta x, \mathcal{H}_{\not\subset})$ to C.
            If $\mathcal{F}_{\mathcal{H}}$ receives $(\text{gcc-ok})$ from C in next round, then $\mathcal{F}_{\mathcal{H}}$ broadcasts $m_{gcc}^{AC} := (\text{gcc}, sid, \beta_{AC}, \beta_{BD}, \Delta x, \mathcal{H}_{\not\subset})$ to both A, C, B, and D then goes to the next phase. 
            Otherwise, $\mathcal{F}_{\mathcal{H}}$ terminates the procedure.
            Note that the interactions with B and D are handled analogously.
    
            \item{\textbf{Capacity transfer phase}} \\
            Upon receiving $(\text{iou}$, $sid$, $\beta_{AC}$, $\beta_{BD}$, $\Delta x$, $\mathcal{H}_{\not\subset}$, $m_{gcc}^{AC}$, $m_{gcc}^{BD})$ from A in round 3 and $(\text{receipt}$, $sid$, $\beta_{AC}$, $\beta_{BD}$, $\Delta x$, $\mathcal{H}_{\not\subset}$, $m_{gcc}^{AC}$, $m_{gcc}^{BD})$ from B in round 4, 
            $\mathcal{F}_{\mathcal{H}}$ updates the channel space $\mathcal{B}_{\not\subset}$ such that $(\beta_{AC} \to \{c - \Delta x, \theta_{w}^{AC}\})$ and $(\beta_{BD} \to \{c + \Delta x, \theta^{BD}_{w}\})$, 
            outputs $m_{\text{ct}} := (\text{transferred}$, $sid$, $\beta_{AC}$, $\beta_{BD}$, $\Delta x$, $m_{gcc}^{AC}$, $m_{gcc}^{BD})$, 
            and proceeds to the next phase. 
            Otherwise, $\mathcal{F}_{\mathcal{H}}$ terminates the procedure.
    
            \item{\textbf{In-channel update phase}} \\
            Upon receiving $(\text{icu-request})$ from A in round 6, 
            $\mathcal{F}_{\mathcal{H}}$ sets $(\beta_{AC} \to \{c - \Delta x, \theta_{w+1}\})$, 
            where $\theta_{w+1}(A) = \theta_w(A) - \Delta x$, $\theta_{w+1}(C) = \theta_w(C)$, 
            and sends $(\text{inChannelUpdate}, sid, \beta_{AC}, m_{\text{ct}}, \theta_{w+1})$ to C. 
            If next round C replies with $(\text{conf-ok})$ then $\mathcal{F}_{\mathcal{H}}$ sends $(\text{confirm}, sid, m_{\text{ct}}, \theta_{w+1})$ to A and outputs $(\text{cc-transferred})$.
            Note that the interactions with B and D are handled analogously.
    
        \end{itemize}
    
        \medskip
        \centerline{\textbf{Withdraw from channel hub}}
    
        Upon receiving $(\text{withdraw}, sid, \beta_{AC}, \mathcal{H}_{\not\subset})$ from A in round 1, $\mathcal{F}_{\mathcal{H}}$ checks if $\beta_{AC} \in \mathcal{B}_{\not\subset}$. If this is not the case then $\mathcal{F}_{\mathcal{H}}$ ignores the withdrawal request. 
        Otherwise, in round 3, $\mathcal{F}_{\mathcal{H}}$ marks channel $\beta_{AC}$ as withdrawn and outputs $(\text{withdrawn})$. 
        The withdrawal can be done in round 2 in the optimistic case, i.e., if $\mathcal{F}_{\mathcal{H}}$ also receives $(\text{withdraw}, sid, \beta_{AC}, \mathcal{H}_{\not\subset})$ from C in round 2.
    
        \medskip
        \centerline{\textbf{Close a payment channel}}
    
        Upon receiving $(\text{close}, sid, \beta_{AC})$ from A in round 1, $\mathcal{F}_{\mathcal{H}}$ checks if $\beta_{AC} \in \mathcal{B}_{\not\subset}$. 
        If this is the case then $\mathcal{F}_{\mathcal{H}}$ ignores the closing request. 
        Otherwise, in round 3, let $\theta$ be the distribution function of channel $\beta_{AC}$, 
        then $\mathcal{F}_{\mathcal{H}}$ sends $\theta(A)$ coins to A and $\theta(C)$ coins to C,
        removes the entry that corresponds to channel $\beta_{AC}$ from $\mathcal{B}$, and outputs $(\text{closed})$.
        The closing process can be finished in round 2 in the optimistic case, i.e., 
        when $\mathcal{F}_{\mathcal{H}}$ also receives $(\text{close}, sid, \beta_{AC})$ from C in round 2.
    
    \end{tcolorbox}
    \caption{Ideal functionality $\mathcal{F}_{\mathcal{H}}$ for Boros}
    \label{FH}
\end{figure*}

\begin{figure*}[ht]
    \begin{tcolorbox}[colback=white, arc=0pt]
    
        \centerline {\textbf{The Contract Functionality $\mathcal{F}_{\mathcal{C}}$}}
    
        \tcblower
    
        \centerline {\textbf{Open a payment channel}}
    
        (1) Upon receiving $(\text{open}, sid, \beta_{AC}, x_a)$ together with $x_a$ coins from A in round 1, $\mathcal{F}_{\mathcal{C}}$ sends a opening request $(\text{opening}, sid, \beta_{AC})$ to C. \\
        (2) If C replies with $(\text{open}, sid, \beta_{AC}, x_c)$ together with $x_c$ coins in round 2, then $\mathcal{F}_{\mathcal{C}}$ outputs $(\text{opened}, sid, \beta_{AC}, c)$, where $c = x_a + x_c$ denotes the funding capacity of channel $\beta_{AC}$. Otherwise, $\mathcal{F}_{\mathcal{C}}$ sends $x_a$ coins back to A's account and outputs $(\text{open-failed})$.
    
        \medskip
        \centerline {\textbf{Join a channel hub}}
    
        (1) Upon receiving $(\text{join}, sid, \beta_{AC}, \mathcal{H}_{\not\subset}, c, \theta_w, \sigma_{A}, \sigma_{C})$ from C in round 2, where $\sigma_{A}$ and $\sigma_{C}$ denote A and C's signature on message $(\text{join}, sid, \beta_{AC}, \mathcal{H}_{\not\subset}, c, \theta_w)$ respectively, then $\mathcal{F}_{\mathcal{C}}$ sends a message $(\text{join}, sid, \beta_{AC}, \mathcal{H}_{\not\subset}, c)$ together with $c$ coins to $\mathcal{F}_{\mathcal{N}}$, obtains the response $(\text{joined})$ or $(\text{join-failed})$, and sends the response to both A and C in round 3.
    
        \medskip
        \centerline {\textbf{Withdraw from channel hub}}
    
        (1) Upon receiving $(\text{withdraw}, sid, \beta_{AC}, \mathcal{H}_{\not\subset}, c, \theta_w)$ from A in round 1, 
        $\mathcal{F}_{\mathcal{C}}$ sends a withdrawal request $(\text{withdrawing})$ to C. \\
        (2) If C replies with $(\text{withdraw}, sid, \beta_{AC}, \mathcal{H}_{\not\subset}, c, \theta_w)$ in next round, 
        then $\mathcal{F}_{\mathcal{C}}$ sends $(\text{withdraw}, sid, \beta_{AC}, \mathcal{H}_{\not\subset}, c)$ to $\mathcal{F}_{\mathcal{N}}$. Otherwise, $\mathcal{F}_{\mathcal{C}}$ sends the same withdrawal request to $\mathcal{F}_{\mathcal{N}}$ in round 3 \\
        (3) If a response $(\text{withdrawn}, sid, \beta_{AC}, \mathcal{H}_{\not\subset}, c)$ together with $c$ coins arrives from $\mathcal{F}_{\mathcal{N}}$, then $\mathcal{F}_{\mathcal{C}}$ outputs $(\text{withdrawn})$ to both A and C in round 3. Otherwise, outputs $(\text{withdraw-failed})$.    
    
        \medskip
        \centerline {\textbf{Close a payment channel}}
    
        (1) Upon receiving message $(\text{close}, sid, \beta_{AC}, c, \theta_{w1})$ from A in round 1, $\mathcal{F}_{\mathcal{C}}$ notifies C of the closing request $(\text{closing})$. \\
        (2) If C replies with $(\text{close}, sid, \beta_{AC}, c, \theta_{w2})$ in the next round, then $\mathcal{F}_{\mathcal{C}}$ sends $\theta_w(A)$ coins to A's account and $\theta_w(C)$ coins to C's account, where $w = max(w1, w2)$. Otherwise, $\mathcal{F}_{\mathcal{C}}$ sends $\theta_{w1}(A)$ coins to A's account and $\theta_{w1}(C)$ coins to C's account in round 3. In both cases, $\mathcal{F}_{\mathcal{C}}$ outputs $(\text{closed})$ in round 3.
    
    \end{tcolorbox}
    \caption{The contract functionality $\mathcal{F}_{\mathcal{C}}$}
    \label{FC}
\end{figure*}

\begin{figure*}[ht]
    \begin{tcolorbox}[colback=white, arc=0pt]
    
        \centerline {\textbf{The Protocol Boros}}
    
        \tcblower
    
        \centerline{\textbf{Open a payment channel}}
        \begin{enumerate}[leftmargin=*]
            \item[A:] Upon receiving message $(\text{open}, sid, \beta_{AC}, x_a)$ from the environment $\mathcal{Z}$ in round 1, A sends a signed message $(\text{open}, sid, \beta_{AC}, x_a)$ together with $x_a$ coins to $\mathcal{F}_{\mathcal{C}}$ and goes to the Common Steps below.
            \item[C:] Upon receiving $(\text{open}, sid, \beta_{AC}, x_c)$ from $\mathcal{Z}$ in round 1, C waits for the opening request $(\text{opening}, sid, \beta_{AC})$ from $\mathcal{F}_{\mathcal{C}}$. If the message does not arrive then C outputs $(\text{open-failed})$ and terminates this procedure. Otherwise, C replies $\mathcal{F}_{\mathcal{C}}$ with $(\text{open}, sid, \beta_{AC}, x_c)$ with $x_c$ coins in round 2 and goes to the Common Steps.
        \end{enumerate}
        Common Steps: If $P \in \{A, C\}$ receives $(\text{opened}, sid, \beta_{AC}, c)$ from $\mathcal{F}_{\mathcal{C}}$ in round 3 then $P$ outputs $(\text{opened})$ and goes idle. If $P$ receives $(\text{open-failed})$ from $\mathcal{F}_{\mathcal{C}}$ then outputs $(\text{open-failed})$ and terminates this procedure.
    
        \medskip
        \centerline{\textbf{In-channel transfer}}
        \begin{enumerate}[leftmargin=*]
            \item[A:] Upon receiving $(\text{update}$, $sid$, $\beta_{AC}$, $\theta)$ from the environment $\mathcal{Z}$ in round 1, A sends a message $m_u := (\text{updating}$, $sid$, $\beta_{AC}$, $\theta_{w+1})$ to C, where $\theta_{w+1} = \theta$ denotes the distribution function with a new version number $w+1$. If A receives $(\text{update-ok}$, $sid$, $\beta_{AC}$, $\theta_{w+1})$ from C in round 3 then he outputs $(\text{updated})$.
            \item[C: ] Upon receiving the updating request $m_u$ from A, C checks whether $\theta_{w+1}(A) + \theta_{w+1}(C)$ is equal to the latest capacity of channel $\beta_{AC}$ that C is aware of. If this does not hold, then C ignores the updating request $m_u$ and terminates this procedure. Otherwise, C sends a message $(\text{update-request})$ to the environment $\mathcal{Z}$. If $\mathcal{Z}$ replies with $(\text{update-ok})$ then C sends $(\text{update-ok}$, $sid$, $\beta_{AC}$, $\theta_{w+1})$ to A and outputs $(\text{updated})$.
        \end{enumerate}
    
        \centerline{\textbf{Close a payment channel}}
        \begin{enumerate}[leftmargin=*]
            \item[A:] Upon receiving $(\text{close}, sid, \beta_{AC})$ from the environment $\mathcal{Z}$ in round 1, A sends a message $(\text{close}, sid, \beta_{AC}, c, \theta_{w_A})$ to $\mathcal{F}_{\mathcal{C}}$, where $c$ and $\theta_{w_A}$ denote the latest funding capacity and distribution function that A is aware of. If A receives $(\text{closed})$ in round 3 then he outputs $(\text{closed})$ and terminates this procedure.
            \item[C:] Upon receiving $(\text{close}, sid, \beta_{AC})$ from $\mathcal{Z}$ in round 1, C waits for the closing request $(\text{closing}, sid, \beta_{AC}, c, \theta_{w_A})$ from $\mathcal{F}_{\mathcal{C}}$. If the message does not arrive then C outputs $(\text{close-failed})$ and terminates this procedure. Otherwise, C replies $\mathcal{F}_{\mathcal{C}}$ with $(\text{close}, sid, \beta_{AC}, c, \theta_{w_C})$ in round 2. If C receives $(\text{closed})$ in round 3 then he outputs $(\text{closed})$ and terminates this procedure.
        \end{enumerate}
    
    \end{tcolorbox}
    
    \caption{Formal protocol Boros Part \MakeUppercase{\romannumeral 1}}
    \label{BorosPart1}
\end{figure*}

\begin{figure*}
    \begin{tcolorbox}[colback=white, arc=0pt]
    
        \centerline {\textbf{The Protocol Boros}}
    
        \tcblower
    
        \centerline{\textbf{Join a channel hub}}
        \begin{enumerate}[leftmargin=*]
            \item[A:] Upon receiving $(\text{join}$, $sid$, $\beta_{AC}$, $\mathcal{H}_{\not\subset})$ from the environment $\mathcal{Z}$ in round 1, A sends $m_{\text{j}}$ := $(\text{join}$, $sid$, $\beta_{AC}$, $\mathcal{H}_{\not\subset}$, $c$, $\theta_w$, $\sigma_{A})$, where $\sigma_{A}$ denotes A's signature on that message, to C and goes to the Common Steps below.
            \item[C:] Upon receiving the joining request $m_{\text{j}}$ from A, C sends a message $(\text{join-request})$ to the environment $\mathcal{Z}$. If $\mathcal{Z}$ replies with $(\text{join-ok})$ then C computes its signature and sends $(\text{join}, sid, \beta_{AC}, \mathcal{H}_{\not\subset}, c, \theta_w, \sigma_{A}, \sigma_{C})$ to $\mathcal{F}_{\mathcal{C}}$ and goes to the Common Steps.
        \end{enumerate}
        Common Steps: If $P \in \{A, C\}$ receives $(\text{joined}, sid, \beta_{AC}, \mathcal{H}_{\not\subset}, c, \theta_w)$ from $\mathcal{F}_{\mathcal{C}}$ in round 3 then $P$ outputs $(\text{joined})$ and goes idle. If $P$ receives $(\text{join-failed})$ from $\mathcal{F}_{\mathcal{C}}$ then outputs $(\text{join-failed})$ and terminates this procedure.

        \medskip
        \centerline{\textbf{Cross-channel transfer}}
    
        \begin{itemize}[leftmargin=*]
            \item{\textbf{Prepare}}
            \begin{enumerate}[leftmargin=*]
                \item[A:] Upon receiving message $(\text{cc-transfer}, sid, \beta_{AC}, \beta_{BD}, \Delta x, \mathcal{H}_{\not\subset})$ from $\mathcal{Z}$ in round 1. If $\Delta x > \theta_w(A)$ where $\theta_w$ denotes the last distribution function that A is aware of, then A ignores this message, otherwise A sends a message $m_{\text{pcc}} := (\text{pcc}, sid, \beta_{AC}, \beta_{BD}, \Delta x, \mathcal{H}_{\not\subset})$ to C.
                \item[C:] Upon receiving $m_{\text{pcc}}$ from A in round 2, C sends a message $(\text{gcc-request}, sid, \beta_{AC}, \beta_{BD}, \Delta x, \mathcal{H}_{\not\subset})$ to the environment $\mathcal{Z}$. If $\mathcal{Z}$ replies with $(\text{gcc-ok})$ then C broadcasts $m_{\text{gcc}}^{AC} := (\text{gcc}, sid, \beta_{AC}, \beta_{BD}, \Delta x, \mathcal{H}_{\not\subset})$ to both A, B, and D. Otherwise, C terminates this procedure.
            \end{enumerate}
            The Behavior of B and D is handled analogously. Note that at the beginning of round 3, both A, C, B, and D should hold $m_{\text{gcc}}^{AC}$ and $m_{\text{gcc}}^{BD}$. 
            Otherwise, they terminate this procedure.
    
            \item{\textbf{Capacity transfer}}
            \begin{enumerate}[leftmargin=*]
                \item[A:] In round 3, A sends message $m_{\text{iou}} := (\text{iou}, sid, \beta_{AC}, \beta_{BD}, \Delta x, \mathcal{H}_{\not\subset}, m_{\text{gcc}}^{AC}, m_{\text{gcc}}^{BD})$ to the ideal functionality $\mathcal{F}_{\mathcal{N}}$.
                If A receives $msg_{ct} := (\text{transferred}, sid, \beta_{AC}, \beta_{BD}, \Delta x, m_{gcc}^{AC}, m_{gcc}^{BD})$ from $\mathcal{F}_{\mathcal{N}}$ in round 5 then A proceeds to the next phase. Otherwise, A terminates this procedure.
                \item[B:] Upon receiving message $m_{\text{iou}}$ from $\mathcal{F}_{\mathcal{N}}$ in round 4, B checks the validity of $m_{\text{iou}}$ and replies with a signed receipt $(\text{receipt}, sid, \beta_{AC}, \beta_{BD}, \Delta x, \mathcal{H}_{\not\subset}, m_{\text{gcc}}^{AC}, m_{\text{gcc}}^{BD})$ to $\mathcal{F}_{\mathcal{N}}$. If B receives $msg_{ct}$ from $\mathcal{F}_{\mathcal{N}}$ in round 5 then B proceeds to the next phase. Otherwise, B terminates this procedure.
            \end{enumerate}
    
            \item{\textbf{In-channel update}}
            \begin{enumerate}[leftmargin=*]
                \item[A:] In round 6, A sends $m_{\text{icu}} := (\text{inChannelUpdate}, sid, \beta_{AC}, m_{\text{ct}}, \theta_{w+1})$ to C where $\theta_{w+1}(A) = \theta_w(A) - \Delta x$ and $\theta_{w+1}(C) = \theta_w(C)$, $\theta_w$ denotes the last distribution function that A is aware of before that cross-channel transfer. If A receives $m_{\text{conf}} := (\text{confirm}, sid, m_{\text{ct}}, \theta_{w+1})$ from C then A outputs $(\text{cc-transferred})$ and goes idle. 
                \item[C:] If C receives $m_{\text{icu}}$ from A in round 7, C sends a message $(\text{conf-request}, sid, m_{\text{ct}}, \theta_{w+1})$ to the environment $\mathcal{Z}$. If $\mathcal{Z}$ replies with $(\text{conf-ok})$ then C sends the confirmation $m_{\text{conf}}$ to A, outputs $(\text{cc-transferred})$, and goes idle.
            \end{enumerate}
            The Behavior of B and D is handled analogously.
        \end{itemize}
        
        \medskip
        \centerline{\textbf{Withdraw from channel hub}}
        \begin{enumerate}[leftmargin=*]
            \item[A:] Upon receiving $(\text{withdraw}, sid, \beta_{AC}, \mathcal{H}_{\not\subset})$ from $\mathcal{Z}$ in round 1, A sends a message $(\text{withdraw}, sid, \beta_{AC}, \mathcal{H}_{\not\subset}, c, \theta_w)$ to $\mathcal{F}_{\mathcal{C}}$ where $c$ and $\theta_w$ denotes the funding capacity and the distribution function of channel $\beta_{AC}$ that A is aware of. Then A goes to the Common Steps.
            \item[C:] Upon receiving $(\text{withdraw}, sid, \beta_{AC}, \mathcal{H}_{\not\subset})$ from $\mathcal{Z}$ in round 1, C waits for the arrival of $(\text{withdrawing}, sid, \beta_{AC}, \mathcal{H}_{\not\subset}, c, \theta_w)$ from $\mathcal{F}_{\mathcal{C}}$. If the message does not arrive then C outputs $(\text{withdraw-failed})$ and terminates this procedure. Otherwise, C checks the validity of $c$ and $\theta_w$, replies with $(\text{withdraw}, sid, \beta_{AC}, \mathcal{H}_{\not\subset}, c, \theta_w)$, and goes to the Common Steps.
        \end{enumerate}
        Common Steps: If $P \in \{A, C\}$ receives $(\text{withdrawn}, sid, \beta_{AC}, \mathcal{H}_{\not\subset}, c, \theta_w)$ from $\mathcal{F}_{\mathcal{C}}$ in round 3 then $P$ outputs $(\text{withdrawn})$ and goes idle. If $P$ receives $(\text{withdraw-failed})$ then outputs $(\text{withdraw-failed})$ and terminates this procedure.
    
    \end{tcolorbox}
    
    \caption{Formal protocol Boros Part \MakeUppercase{\romannumeral 2}}
    \label{BorosPart2}
\end{figure*}

The ideal functionality $\mathcal{F}_{\mathcal{H}}$, as shown in Figure \ref{FH}, maintains two channel spaces.
One of the two channel space consists of payment channels that have not yet joined the channel hub. 
We denote it as $\mathcal{B} : \beta \to \{c, \theta_w\}$, where $\beta \in \{0, 1\}^\lambda$ denotes the contract account of channel $\beta$, $c \in \mathbb{N}_0$ denotes the funding capacity of channel $\beta$, and $\theta_w$ denotes the distribution function of its total funds corresponds to the version number $w$. 
For example, in channel $\beta_{AC}$, we use $\theta_w^{AC}(A)$ to denotes A's balance and $\theta_w^{AC}(C)$ to denote C's balance corresponds to the version number $w$.
We note that when it does not affect the clarity of expression, we often omit the superscript.
Moreover, we always have $c = \theta_w(A) + \theta_w(C)$ and $\theta_w(P_i) \geq 0$. 
The monotonically increasing version number $w$ is used for tracing every transfer in a payment channel and is initially set to 1. 
The other channel space, which is denoted as $\mathcal{B}_{\not\subset} : \beta \to \{c, \theta_w\}$, consists of payment channels that have already joined the channel hub.
When we say that a payment channel $\beta$ is marked as joined, we mean that the channel $\beta$ has been moved from channel space $\mathcal{B}$ to $\mathcal{B}_{\not\subset}$.
Similarly, a payment channel $\beta$ is marked as withdrawn means that the channel $\beta$ has been moved from $\mathcal{B}_{\not\subset}$ back to $\mathcal{B}$. 

The ideal functionality $\mathcal{F}_{\mathcal{H}}$ offers the following interfaces for the parties: 
(1) \textit{open a payment channel} between two parties. 
When receiving both the opening requests from A and C together with $x_a$ and $x_c$ coins respectively, a payment channel $\beta_{AC}$ of funding capacity $c = x_a + x_c$ is opened.
(2) \textit{in-channel transfer}. 
A two-phase process for performing off-chain transfers. 
When triggered by A with an update request, $\mathcal{F}_{\mathcal{H}}$ asks C for the confirmation of that transfer. 
Once C replies with his confirmation, $\mathcal{F}_{\mathcal{H}}$ updates the distribution function for channel $\beta_{AC}$ and outputs $(\text{updated})$.
(3) \textit{join a channel hub.} 
A payment channel can join a channel hub only when both parties reach an agreement and the payment channel has not yet joined any channel hub. 
When triggered by A with a joining request, $\mathcal{F}_{\mathcal{H}}$ asks C for the confirmation of that joining.
Once C replies with his confirmation, $\mathcal{F}_{\mathcal{H}}$ marks $\beta_{AC}$ as joined and notifies about the result through the message (joined).
(4) \textit{cross-channel transfers.} 
It can be performed only between two payment channels that have already joined in the channel hub. 
This process is divided into three phases.
In the prepare phase, both channel $\beta_{AC}$ and $\beta_{BD}$ should reach an agreement on this transfer.
In the capacity transfer phase, if $\mathcal{F}_{\mathcal{H}}$ receives both the iou message from A and the receipt from B, then $\mathcal{F}_{\mathcal{H}}$ updates its internal state such that the funding capacity of channel $\beta_{AC}$ decreases by $\Delta x$ coins and $\beta_{BD}$ increases by the same amount. We note that at the end of this phase, the distribution function of both channel $\beta_{AC}$ and $\beta_{BD}$ remains unchanged.
The last phase is triggered by in-channel update requests from A and B respectively. $\mathcal{F}_{\mathcal{H}}$ then asks C and D for their confirmation, which finally results in the changing of the distribution function of channel $\beta_{AC}$ and $\beta_{BD}$.
(5) \textit{withdraw from channel hub.} 
A payment channel can withdraw from the channel hub only when the payment channel has already joined in the channel hub. 
When receiving the withdrawal request from A or C, $\mathcal{F}_{\mathcal{H}}$ finally marks channel $\beta_{AC}$ as withdrawn and outputs $(\text{withdrawn})$. 
Our ideal functionality guarantees that an honest party will always manage to withdraw his payment channel from the channel hub in a fixed time.
(6) \textit{close a payment channel.} 
When triggered by a closing request from A or C, $\mathcal{F}_{\mathcal{H}}$ finally refunds A with $\theta(A)$ coins and C with $\theta(C)$ coins within three rounds. 
Our ideal functionality guarantees that an honest party will always manage to close the payment channel and get refunded in a fixed time.

Now we discuss how our ideal functionality $\mathcal{F}_{\mathcal{H}}$ satisfies the security properties mentioned in Section \ref{securityProperties}.
\begin{itemize}[leftmargin=*]
    \item \textbf{Consensus on channel hub enrollment and withdrawal.} 
        Once a payment channel joins in or withdraws from a channel hub, the ideal functionality $\mathcal{F}_{\mathcal{H}}$ would notify all parties of the results through messages (joined), (withdrawn) or (join-failed). Thus it is straightforward to see that this property always holds.
    \item \textbf{Consensus on channel capacity.}
        The initial funding capacity of a payment channel is settled and notified by the ideal functionality $\mathcal{F}_{\mathcal{H}}$ when successfully opening the channel. 
        Besides, an honest party is guaranteed to be notified by the $\mathcal{F}_{\mathcal{H}}$ of the results of capacity transfers through message $m_{\text{ct}}$.
        Thus, an honest party can always learn the funding capacity of its payment channels.
    \item \textbf{Balance security.}
        The analysis of balance security consists of two points. One is the consensus on the funding capacity of the payment channel. The other is the consensus on the final distribution function of the payment channel. 
        The former is discussed above. 
        For the distribution function, $\mathcal{F}_{\mathcal{H}}$ always guarantees that the distribution function with larger version number $w$ always wins when closing the payment channel.
        Thus, an honest party with the latest distribution function is guaranteed to be paid out with the correct amount of coins when closing a payment channel.
\end{itemize}




\subsection{The Boros Protocol} \label{theBorosProtocol}

Now we formally describe the Boros protocol, which consists of the contract functionality $\mathcal{F}_{\mathcal{C}}$ as shown in Figure \ref{FC} and the specification of the behavior of all involved parties as shown in Figure \ref{BorosPart1} and Figure \ref{BorosPart2}.

We firstly discuss those operations that share with traditional payment channels, which are shown in Figure \ref{BorosPart1}. 
To open a payment channel, A deploys a new instance of contract $\mathcal{F}_\mathcal{C}$ together with $x_a$ coins. 
Upon construction, the contract $\mathcal{F}_\mathcal{C}$ notifies C with the opening request. 
Once the contract gets a confirmation from C together with $x_c$ coins in round 2, then the payment channel $\beta_{AC}$ is opened. 
Otherwise, the contract $\mathcal{F}_\mathcal{C}$ refunds A with $x_a$ coins and outputs $(\text{open-failed})$.

Once the payment channel $\beta_{AC}$ is opened, A and C can perform in-channel transfers without involving the blockchain. 
Firstly, A sends an update request to C, which contains a new distribution function $\theta_{w+1}$. 
When C receives that request, it should check the validity of the capacity of channel $\beta_{AC}$, that is, $\theta_{w+1}(A) + \theta_{w+1}(C)$ should always be equal to the latest capacity of channel $\beta_{AC}$.
If this is the case, then C sends a message $(\text{update-request})$ to the environment $\mathcal{Z}$, asking if it agrees for an update. 
If the environment responds with $(\text{update-ok})$, then C sends A with a confirmation to complete that in-channel transfer. 
We emphasize that the validity check of channel capacity is necessary since the funding capacity of our payment channel could be changed by cross-channel transfers.

To close a payment channel $\beta_{AC}$, A sends a closing request containing his last distribution function $\theta_{w1}$ to the contract instance $\mathcal{F}_\mathcal{C}$. 
Upon receiving the closing request from A, the contract then forwards that request to C. 
If next round C replies with his last distribution function $\theta_{w2}$, 
then the contract $\mathcal{F}_\mathcal{C}$ chooses the latest one, 
which is denoted as $\theta_w$ where $w = max(w1, w2)$, 
and sends coins back to accounts of both A and C according to $\theta_w$. 
Otherwise, the contract $\mathcal{F}_\mathcal{C}$ sends coins back to accounts of both A and C according to $\theta_{w1}$.

Then we discuss those extended operations, which are shown in Figure \ref{BorosPart2}.
We start with the \textit{joining} process. 
To join a channel hub, A sends a message containing the target channel hub instance $\mathcal{H}_{\not\subset}$, the funding capacity $c$, the latest distribution function $\theta_w$ of channel $\beta_{AC}$, and A's signature $\sigma_A$ on that message to C.
Upon receiving the joining request from A, C sends a message $(\text{join-request})$ to the environment $\mathcal{Z}$, asking if it agrees for that joining.
If the environment responds with $(\text{join-ok})$, then C sends the joining request containing both A and C's signature to the contract $\mathcal{F}_\mathcal{C}$.
Once the contract receives that signed joining request from C, it sends a joining request together with $c$ coins to the channel hub $\mathcal{F}_{\mathcal{N}}$ to finish the joining process. 
Otherwise, the joining procedure is considered failed.

Now we discuss the \textit{cross-channel transfer} procedure, which is divided into three phases.
In the first phase, A sends a message $m_\text{pcc}$ to C, 
indicating his intention to perform a cross-channel transfer, 
which moves $\Delta x$ coins using A's balance in channel $\beta_{AC}$ to channel $\beta_{BD}$ via the channel hub $\mathcal{H}_{\not\subset}$. 
When C receives $m_\text{pcc}$ from A, he checks if the A's balance in channel $\beta_{AC}$ exceeds $\Delta x$. 
If this is not the case, then C rejects the cross-channel transfer request.
Otherwise, if C agrees with that transfer, then he attaches his signature on the $m_\text{pcc}$ to produce $m_\text{gcc}$, which indicates C's permission on that cross-channel transfer. 
Then C broadcasts the message $m_\text{gcc}$ to A, B, and D, and goes to the third phase. 
The interaction between B and D is handled analogously. 
In the capacity transfer phase, the initiator A sends an iou message to the channel hub $\mathcal{H}_{\not\subset}$. 
As mentioned in the channel hub functionality $\mathcal{F}_{\mathcal{N}}$, the channel hub $\mathcal{H}_{\not\subset}$ forwards the iou request to B, obtains B's signed receipt, and then executes that transfer. 
The execution will cause the funding capacity of channel $\beta_{AC}$ to be decreased by $\Delta x$ coins and $\beta_{BD}$ increased by the same amount. 
If the execution successes, then $\mathcal{H}_{\not\subset}$ will notify both A and B about execution result through message $m_{\text{ct}}$.
In the in-channel update phase, A sends the result of the second phase to C. 
The message sent by A should include a new distribution function $\theta_{w+1}$ such that $\theta_{w+1}(A) = \theta_w(A) - \Delta x$ and $\theta_{w+1}(C) = \theta_w(C)$, and the signed result from the operator server of channel hub $\mathcal{H}_{\not\subset}$. 
Upon receiving this message, C checks its validity and replies with a confirmation to finish the cross-channel transfer.

Finally, we discuss the \textit{withdrawal} procedure.
To withdraw from the channel hub, A sends a message containing the latest funding capacity $c$ and distribution function $\theta_w$ of channel $\beta_{AC}$ to the contract $\mathcal{F}_\mathcal{C}$. 
Upon receiving the withdrawal request from A, the contract $\mathcal{F}_\mathcal{C}$ notifies to C of that withdrawal request. 
Once receiving a signed reply from C, the contract $\mathcal{F}_\mathcal{C}$ sends a withdrawal request to the channel hub $\mathcal{F}_{\mathcal{N}}$ to finish the withdrawal process.

\subsection{Security Definition}

Now we formally state the security of our Boros protocol.
Due to the page limit, formal security proof is provided in Appendix A.

\begin{theorem}
\label{the}
    Protocol Boros securely realizes functionality $\mathcal{F}_{\mathcal{H}}$ in the $(\mathcal{F}_{\mathcal{L}}, \mathcal{F}_{\mathcal{N}}, \mathcal{F}_{\mathcal{C}})\text{-}hybrid$ model.
\end{theorem}


\begin{table}[ht]
    \renewcommand{\arraystretch}{1.3}
    \caption{The Execution Cost of Each Operation.}
    \label{t1}
    \begin{tabular}{|l|c|ccc|c|c|}
        \hline
           & \makecell{\#\\on-\\chain} & \multicolumn{3}{c|}{cost} & \makecell{\#\\off-\\chain} & \# sigs \\
            & txs & gas & Ether & USD &  msgs & \\
        \hline
        open & 2 & 173147 & 0.0034 & 0.84 & 0 & 2 \\
        \hline
        in-channel transfer & 0 & 0 & 0 & 0 & 2 & 2 \\
        \hline
        join & 1 & 154723 & 0.0030 & 0.75 & 0 & 2 \\
        \hline
        cross-channel transfer & 0 & 0 & 0 & 0 & 17 & 17 \\
        \hline
        withdraw & 2 & 97749 & 0.0019 & 0.47 & 0 & 2 \\
        \hline
        close & 2 & 148413 & 0.0029 & 0.72 & 0 & 2 \\
        \hline
    \end{tabular}
\end{table}

\section{Implementation And Evaluation} \label{implementation}



We implement the Boros protocol on Ethereum and measure the execution cost of each operation in the Boros protocol. Note that our current implementation aims at demonstrating the feasibility of the Boros protocol and we will further optimize it in future work. 
Moreover, we simulate off-chain payment networks of  different sizes and prove that our construction can effectively shorten the average transaction path length.

Ethereum uses gas to measure the amount of computational resource used to execute certain operations.
Every instruction executed by the Ethereum Virtual Machine (EVM) costs a certain amount of gas.
There is no fixed price of the conversion between gas and Ether. 
It is up to the sender of a transaction to specify a gas price, 
which affects the willingness of the miners to process the transaction.
A lower gas price results in longer waiting time for a transaction to be mined.
The average gas price is typically on the order of about 20 Gwei (or $2 \times 10^{-8}$ Ether). 
When we prepar this paper, the exchange rate of the Ether against the US dollar is 1:243.2. That is, $1 \text{ gas} = 2 \times 10^{-8} \text{ Ether} = 4.864 \times 10^{-6} \text{ USD}$.

Our evaluation adopts the following criteria: 
the number of on-chain transactions, 
the execution cost measured in gas, Ether, and USD, 
the number of off-chain messages, 
and the number of signatures, 
which is dominant for both message length and computational complexity.
Table \ref{t1} shows the execution cost of each operation under these metrics.

\begin{table*}[t]
    \renewcommand{\arraystretch}{1.3}
    \caption{Avg. Path Length With One Hub}
    \label{t2}
    \centering
    \rowcolors{2}{gray!15}{}
    \begin{tabular}{cccccccccccc}
        \hline
        $\alpha$ & \# of Nodes & PN-FW & PH-FW & CH-FW & $\Delta1$-FW & $\Delta2$-FW & PN-SM & PH-SM & CH-SM & $\Delta1$-SM & $\Delta2$-SM \\ \hline
        
        & 200 & 2.77 & 2.74 & 2.66 & 3.0\% & 3.9\% & 4.01 & 3.99 & 3.65 & 8.7\% & 9.1\% \\
        \cellcolor{white} & 400 & 3.10 & 3.03 & 2.93 & 3.4\% & 5.5\% & 4.71 & 4.58 & 4.03 & 11.9\% & 14.5\% \\
        & 600 & 3.30 & 3.19 & 3.05 & 4.5\% & 7.4\% & 5.06 & 4.76 & 4.32 & 9.3\% & 14.5\% \\
        \cellcolor{white} & 800 & 3.44 & 3.32 & 3.12 & 5.9\% & 9.4\% & 5.33 & 4.92 & 4.27 & 13.3\% & 20.0\% \\
        \multirow{-5}*{5\%} 
        & 1000 & 3.55 & 3.40 & 3.22 & 5.5\% & 9.3\% & 5.53 & 5.13 & 4.20 & 18.3\% & 24.2\% \\ \hline
        
        \cellcolor{white} & 200 & 2.76 & 2.67 & 2.50 & 6.2\% & 9.5\% & 4.04 & 3.88 & 3.31 & 14.7\% & 18.1\% \\
        & 400 & 3.09 & 2.90 & 2.70 & 6.8\% & 12.7\% & 4.59 & 4.04 & 3.53 & 12.6\% & 23.0\% \\
        \cellcolor{white} & 600 & 3.29 & 3.04 & 2.78 & 8.5\% & 15.6\% & 5.02 & 4.31 & 3.49 & 19.1\% & 30.5\% \\
        & 800 & 3.42 & 3.12 & 2.81 & 10.0\% & 17.8\% & 5.25 & 4.40 & 3.54 & 19.5\% & 32.6\% \\
        \multirow{-5}*{10\%} 
        \cellcolor{white} & 1000 & 3.53 & 3.19 & 2.86 & 10.4\% & 19.1\% & 5.58 & 4.28 & 3.48 & 18.7\% & 37.5\% \\ \hline
        
        & 200 & 2.74 & 2.57 & 2.28 & 11.3\% & 16.8\% & 4.01 & 3.53 & 2.84 & 19.6\% & 29.2\% \\
        \cellcolor{white} & 400 & 3.08 & 2.75 & 2.48 & 9.8\% & 19.4\% & 4.62 & 3.75 & 2.99 & 20.2\% & 35.3\% \\
        & 600 & 3.26 & 2.84 & 2.50 & 12.0\% & 23.4\% & 4.96 & 3.72 & 2.97 & 20.1\% & 40.0\% \\
        \cellcolor{white} & 800 & 3.41 & 2.88 & 2.53 & 12.2\% & 25.6\% & 5.31 & 3.40 & 2.92 & 14.1\% & 45.0\% \\
        \multirow{-5}*{15\%} 
        & 1000 & 3.50 & 2.92 & 2.56 & 12.4\% & 26.9\% & 5.49 & 3.62 & 2.71 & 25.2\% & 50.7\% \\ \hline
     \end{tabular}
\end{table*}

\begin{table*}[t]
    \renewcommand{\arraystretch}{1.3}
    \caption{Avg. Path Length With Multiple Hubs}
    \label{t3}
    \centering
    \begin{tabular}{cccccccccccccc}
        \hline
        $\alpha$ & \# of Nodes & Hub Size & \# of Hubs & PN-FW & PH-FW & CH-FW & $\Delta1$-FW & $\Delta2$-FW & PN-SM & PH-SM & CH-SM & $\Delta1$-SM & $\Delta2$-SM \\ \hline
        
        & 5000 & 100 & 3 & 4.31 & 4.03 & 3.72 & 7.7\% & 13.7\% & 7.06 & 5.98 & 4.95 & 17.2\% & 29.9\% \\ \rowcolor{gray!15}
        \cellcolor{white} & 5000 & 200 & 2 & 4.31 & 3.95 & 3.64 & 7.8\% & 15.7\% & 7.09 & 5.60 & 4.50 & 19.7\% & 36.5\% \\ \cline{2-14}
        & 10000 & 100 & 5 & 4.64 & 4.30 & 3.93 & 8.8\% & 15.4\% & 7.77 & 6.47 & 5.14 & 20.6\% & 33.9\% \\ \rowcolor{gray!15}
        \multirow{-4}*{5\%}
        \cellcolor{white} & 10000 & 200 & 3 & 4.64 & 4.20 & 3.81 & 9.4\% & 17.9\% & 7.76 & 6.06 & 4.70 & 22.4\% & 39.4\% \\ \hline
        
        & 5000 & 100 & 5 & 4.30 & 3.79 & 3.35 & 11.5\% & 21.9\% & 7.07 & 5.25 & 4.16 & 20.7\% & 41.2\% \\ \rowcolor{gray!15}
        \cellcolor{white} & 5000 & 200 & 3 & 4.29 & 3.67 & 3.22 & 12.3\% & 25.0\% & 7.08 & 4.83 & 3.79 & 21.4\% & 46.4\% \\ \cline{2-14}
        & 10000 & 100 & 10 & 4.61 & 4.07 & 3.56 & 12.6\% & 22.9\% & 7.70 & 5.91 & 4.41 & 25.5\% & 42.8\% \\ \rowcolor{gray!15}
        \multirow{-4}*{10\%}
        \cellcolor{white} & 10000 & 200 & 5 & 4.62 & 3.90 & 3.40 & 12.8\% & 26.3\% & 7.74 & 5.24 & 4.00 & 23.6\% & 48.3\% \\ \hline
        
        & 5000 & 100 & 8 & 4.27 & 3.64 & 3.14 & 13.8\% & 26.4\% & 7.01 & 5.14 & 3.83 & 25.5\% & 45.4\% \\ \rowcolor{gray!15}
        \cellcolor{white} & 5000 & 200 & 4 & 4.27 & 3.47 & 2.99 & 14.0\% & 30.0\% & 7.02 & 4.47 & 3.43 & 23.3\% & 51.2\% \\ \cline{2-14}
        & 10000 & 100 & 15 & 4.59 & 3.90 & 3.32 & 14.9\% & 27.8\% & 7.70 & 5.65 & 4.03 & 28.5\% & 47.6\% \\ \rowcolor{gray!15}
        \multirow{-4}*{15\%}
        \cellcolor{white} & 10000 & 200 & 8 & 4.59 & 3.71 & 3.17 & 14.7\% & 31.1\% & 7.70 & 5.03 & 3.74 & 25.6\% & 51.4\% \\ \hline
    \end{tabular}
\end{table*}

Then, we simulate payment networks of different sizes to evaluate the effectiveness of our construction on shortening the average path length.
In particular, we control the overall construction cost and then measure the average path length for each approach.
We first set up an underlying payment network of different size, then
(a) open additional payment channels 
(b) set up one/multiple payment hub
(c) set up one/multiple channel hub.
We ensure that the above settings cost the same with each other.
For comparison with bare payment network,
since the execution cost of a join operation is similar to opening a new payment channel in terms of gas cost (see Table \ref{t1}),
we randomly open the same number of payment channels on the underlying payment network.
Then we extract transactions from the Ripple dataset \cite{rippleDataSet} and replay them in those settings.
To find a path for each transaction, we use the SpeedyMurmurs \cite{roos2018settling} algorithm and the Floyd-Warshall shortest path algorithm respectively.

For the case with only one channel hub, 
since a payment/channel hub cannot hold too many participants (limited by the operator server), 
we set up small payment networks where the number of nodes ranges from 200 to 1000.
An important system parameter is the ratio of the number of nodes to payment channels, which determines the density of the payment network.
In fact, the higher the density of the payment network, the lower the average path length between nodes, and the less necessary to deploy the channel hub.
We refer to the Ripple dataset \cite{rippleDataSet} and the Lightning Network \cite{1ML}.
When we prepare this article, the Ripple data set contains 67149 nodes and 199574 edges with a ratio of 2.97,
the Lightning Network contains 8655 nodes and 34696 edges with a ratio of 4.0.
We choose 4 as the ratio of the number of nodes to payment channels of the payment network for our simulations.
Another important system parameter is the joining ratio, denoted as $\alpha$, which indicates how many participants will be joined into the payment/channel hub.
Table \ref{t2} shows the simulation results with only one channel hub, where
PN denotes the ``Payment Network'', 
PH denotes the ``Payment Hub'', 
CH denotes the ``Channel Hub'',
FW denotes the Floyd-Warshall shortest path algorithm,
and SM denotes the SpeedyMurmurs routing algorithm.
$\Delta1$ corresponds to the improvement of our construction over the payment hub,
and $\Delta2$ corresponds to the promotion over the payment network.

For the case of multiple channel hubs, the size of the underlying payment network can be further expanded. 
We simulate payment networks of 5,000 nodes and 10,000 nodes.
In these cases, we limit the maximum number of participants per hub, denoted as $k$, ranging from 100 to 200, and then create multiple payment/channel hubs.
The number of hubs can be easily calculated by $\lceil \frac{n \times \alpha}{k} \rceil$, where $n$ denotes the size of the payment network.
Similarly, we set the ratio of the number of nodes to payment channels of the underlying payment network to 4 and test the effect of different $\alpha$ and $k$ on the average path length.
Table 2 shows the simulation results.

\section{Conclusion} \label{conclusion}

In this paper we propose channel hub to support transferring coints directly from one payment channel to another within the same hub. Base on this idea, we design a novel protocol named Boros to perform secure off-chain cross-channel transfers.
The Boros protocol guarantees that an honest party will not bear any financial losses despite strong adversarial capabilities.
We present the security definition of the Boros protocol formally and prove its security using the UC-framework.
Moreover, we develop a prototype on Ethereum and measure the execution cost of each operation in the Boros protocol. Our evaluation on payment networks of different configurations shows that our protocol can effectively shorten the off-chain routing path.
In future work, we will investigate how to optimize the channel hub. For example, in current design, all transactions in a channel hub are fully ordered and executed by the operator server, but a considerable number of transactions can be partially ordered. Exploiting this observation may improve the performance.







%

\bibliographystyle{IEEEtranS}
\bibliography{refList}

\appendices


\section{Security Proof} \label{securityProof}


\begin{IEEEproof}{Proof of Theorem \ref{the}}

Let $\mathcal{A}$ be an adversary that interacts with parties running protocol Boros in the $(\mathcal{F}_{\mathcal{L}}, \mathcal{F}_{\mathcal{N}}, \mathcal{F}_{\mathcal{C}})$-hybrid model. 
We construct an ideal-process adversary $\mathcal{S}$, which is also called the simulator, such that the view of any environment $\mathcal{Z}$ from an interaction with $\mathcal{A}$ and Boros is distributed identically to its view of interaction with $\mathcal{S}$ in the ideal process for $\mathcal{F}_{\mathcal{H}}$.
As usual, the simulator $\mathcal{S}$ runs a simulated copy of $\mathcal{A}$.
Any input from environment $\mathcal{Z}$ is forwarded to $\mathcal{A}$, and any output of $\mathcal{A}$ is copied to the output of $\mathcal{S}$.

In addition, $\mathcal{S}$ proceeds as follows:

\subsubsection*{Open a payment channel}
Simulating the opening procedure is quite straightforward. 
We let $\mathcal{S}$ simulate the behaviors of the contract functionality $\mathcal{F}_{\mathcal{C}}$ and honest parties.
Then opening procedure starts from $\mathcal{Z}$ sending $(\text{open}$, $sid$, $\beta_{AC}$, $x_a)$ to A and $(\text{open}$, $sid$, $\beta_{AC}$, $x_c)$ to C. 
If A is not corrupted and sends a message $(\text{open}$, $sid$, $\beta_{AC}$, $x_a)$ together with $x_a$ coins to $\mathcal{F}_{\mathcal{C}}$ in round 1, 
then $\mathcal{S}$ sends $(\text{transfer}$, $sid$, $A$, $\alpha_{\mathcal{C}}$, $x_a)$ to the ledger functionality $\mathcal{F}_{\mathcal{L}}$, and $(\text{opening}$, $sid$, $\beta_{AC})$ to C in the name of $\mathcal{F}_{\mathcal{C}}$. 
Otherwise, if A is corrupted, then $\mathcal{S}$ terminates the simulation.
Next round if C sends $(\text{open}$, $sid$, $\beta_{AC}$, $x_c)$ together with $x_c$ coins to $\mathcal{F}_{\mathcal{C}}$, 
then $\mathcal{S}$ sends $(\text{transfer}$, $sid$, $C$, $\alpha_{\mathcal{C}}$, $x_c)$ to $\mathcal{F}_{\mathcal{L}}$, and outputs $(\text{opened}$, $sid$, $\beta_{AC}$, $c)$ in the name of $\mathcal{F}_{\mathcal{C}}$.
Otherwise, if C is corrupted and does not send his $(\text{open})$, then $\mathcal{S}$ sends $(\text{transfer}$, $si$d,$ \alpha_{\mathcal{C}}$, $A$, $x_a)$ to $\mathcal{F}_{\mathcal{L}}$ in the name of $\mathcal{F}_{\mathcal{C}}$ and outputs $(\text{open-failed})$ in round 3.

\subsubsection*{In-channel transfer}
To simulate the in-channel transfer procedure, we just let $\mathcal{S}$ simulate the behaviors of honest parties.
The in-channel transfer procedure starts from $\mathcal{Z}$ sending $(\text{update}$, $sid$, $\beta_{AC}$, $\theta)$ to the initiator A.
If A sends $(\text{updating}$, $sid$, $\beta_{AC}$, $\theta)$ to C, then $\mathcal{S}$ sends $(\text{update}$, $sid$, $\beta_{AC}$, $\theta)$ to the ideal functionality $\mathcal{F}_{\mathcal{H}}$ in the name of A.
If C is not corrupted and sends $(\text{update-ok}$, $sid$, $\beta_{AC}$, $\theta)$ to A in next round, then $\mathcal{S}$ sends $(\text{update-ok}$, $sid$, $\beta_{AC}$, $\theta)$ to $\mathcal{F}_{\mathcal{H}}$ in the name of C.
If C is corrupted and does not send $(\text{update-ok}$, $sid$, $\beta_{AC}$, $\theta)$ to A, then A fails to learn the result of that update. 
According to our assumption (mentioned in Section \ref{informalDescription}), an honest party does not engage in any other further transfers if he detects that some other party is dishonest.
In such a situation, the honest party will terminate the protocol with the dishonest one, which means that he will launch a withdrawal from channel hub if needed and then close the payment channel.
One possible case is that the update is beneficial to C and C is rational. 
In such case, C will commit the result of this update to the contract functionality $\mathcal{F}_{\mathcal{C}}$ when closing the channel. 
In other words, this update does happen no matter if C does not send the $(\text{update-ok}$, $sid$, $\beta_{AC}$, $\theta$) to A. 
Thus we simply let $\mathcal{S}$ send a $(\text{update-ok}$, $sid$, $\beta_{AC}$, $\theta)$ to $\mathcal{F}_{\mathcal{H}}$ in the name of C to make this update happened.
In other cases, either when this update is not beneficial to C or C is irrational, C does not commit the result of this update when closing the channel.
In these cases, we just let $\mathcal{S}$ move the appropriate amount of funds from C to A using the \textit{transfer} interface provided by $\mathcal{F}_{\mathcal{L}}$.

\subsubsection*{Join a channel hub}
To simulate the joining procedure, we just let $\mathcal{S}$ simulate the behaviors of honest parties, contract functionality $\mathcal{F}_{\mathcal{C}}$ and channel hub functionality $\mathcal{F}_{\mathcal{N}}$. 
Note that when successfully joined, $\mathcal{S}$ has to simulate the transfer of funding capacity $c$ from the special account of $\mathcal{F}_{\mathcal{C}}$ (denoted as $\alpha_{\mathcal{C}}$) to $\mathcal{F}_{\mathcal{N}}$ (denoted as $\alpha_{\mathcal{N}}$).
The joining procedure starts from $\mathcal{Z}$ sending $(\text{join}$, $sid$, $\beta_{AC}$, $\mathcal{H}_{\not\subset})$ to A. 
If A is not corrupted and sends a message $(\text{join}$, $sid$, $\beta_{AC}$, $\mathcal{H}_{\not\subset}$, $c$, $\theta_w, \sigma_A)$ to C in round 1, then $\mathcal{S}$ sends $(\text{join}, sid, \beta_{AC}, \mathcal{H}_{\not\subset})$ to the ideal functionality $\mathcal{F}_{\mathcal{H}}$ in the name of A. 
Otherwise, if A is corrupted, then $\mathcal{S}$ terminates the simulation. 
Next round if C sends $(\text{join}$, $sid$, $\beta_{AC}$, $\mathcal{H}_{\not\subset}$, $c$, $\theta_w, \sigma_A, \sigma_C)$ to $\mathcal{F}_{\mathcal{C}}$, 
then $\mathcal{S}$ sends $(\text{join}$, $sid$, $\beta_{AC}$, $c)$, where $c$ denotes the funding capacity of channel $\beta_{AC}$$, $to $\mathcal{F}_{\mathcal{N}}$ in the name of $\mathcal{F}_{\mathcal{C}}$,
sends $(\text{transfer}$, $sid$, $\alpha_{\mathcal{C}}$, $\alpha_{\mathcal{N}}$, $c)$ to $\mathcal{F}_{\mathcal{L}}$ in the name of $\mathcal{F}_{\mathcal{N}}$,
 and outputs $(\text{joined})$.

\subsubsection*{Cross-channel transfer} 
To simulate the cross-channel transfer procedure, 
we let $\mathcal{S}$ simulate the behaviors of honest parties and the channel hub functionality $\mathcal{F}_{\mathcal{N}}$. 
The simulation of cross-channel transferring can be divided into three phases. 

In the first phase, namely the prepare phase, A, C, B, and D should reach an agreement on the transfer, which means that all of them should hold the $m_{\text{gcc}}^{AC}$ and $m_{\text{gcc}}^{BD}$ at the end of this phase. 
Otherwise, they abort this transfer. 
The prepare phase starts from $\mathcal{Z}$ sending $(\text{cc-transfer}$, $sid$, $\beta_{AC}$, $\beta_{BD}$, $\Delta x$, $\mathcal{H}_{\not\subset})$ to both A and B.
The interactions between A and C, B and D is analogous. Thus we only take A and C as an example. 
If A sends $(\text{pcc}$, $sid$, $\beta_{AC}$, $\beta_{BD}$, $\Delta x$, $\mathcal{H}_{\not\subset})$ to C in round 1, 
then $\mathcal{S}$ sends $(\text{cc-transfer}$, $sid$, $\beta_{AC}$, $\beta_{BD}$, $\Delta x$, $\mathcal{H}_{\not\subset})$ to $\mathcal{F}_{\mathcal{H}}$ in the name of A.
Otherwise, $\mathcal{S}$ terminates this simulation.
If next round C broadcasts $(\text{gcc}$, $sid$, $\beta_{AC}$, $\beta_{BD}$, $\Delta x$, $\mathcal{H}_{\not\subset})$, then $\mathcal{S}$ sends $(\text{gcc-ok})$ to $\mathcal{F}_{\mathcal{H}}$ in the name of C. 
Otherwise, if C is malicious and does not agree with this transfer, then $\mathcal{S}$ terminates this simulation. 

In the second phase, the capacity transfer is performed through the ideal functionality $\mathcal{F}_{\mathcal{N}}$ by A and B.
Thus we simply let $\mathcal{S}$ simulate the behaviors of functionality $\mathcal{F}_{\mathcal{N}}$ and honest parties. 
If A sends $(\text{iou}$, $sid$, $\beta_{AC}$, $\beta_{BD}$, $\Delta x$, $\mathcal{H}_{\not\subset}$, $m_{\text{gcc}}^{AC}$, $m_{\text{gcc}}^{BD})$ to $\mathcal{F}_{\mathcal{N}}$ in round 3, 
then $\mathcal{S}$ forwards that $(\text{iou})$ request to B in the name of $\mathcal{F}_{\mathcal{N}}$.
Otherwise, $\mathcal{S}$ terminates this simulation.
If next round B replies with $(\text{receipt}$, $sid$, $\beta_{AC}$, $\beta_{BD}$, $\Delta x$, $\mathcal{H}_{\not\subset}$, $m_{\text{gcc}}^{AC}$, $m_{\text{gcc}}^{BD})$ to $\mathcal{F}_{\mathcal{N}}$, 
then $\mathcal{S}$ sends $(\text{transferred}$, $sid$, $\beta_{AC}$, $\beta_{BD}$, $\Delta x$, $\mathcal{H}_{\not\subset}$, $m_{\text{gcc}}^{AC}$, $m_{\text{gcc}}^{BD})$ to both A, C, B and D in the name of $\mathcal{F}_{\mathcal{N}}$ and proceeds to the last phase. 

In the last phase, again, the interaction between A and C, B and D is analogous. Thus we only take A and C as an example.
If A sends $(\text{inChannelUpdate}$, $sid$, $\beta_{AC}$, $m_{\text{ct}}$, $\theta_{w+1})$ to C, then $\mathcal{S}$ sends $(\text{icu-request})$ to $\mathcal{F}_{\mathcal{H}}$ in the name of A. 
If next round C replies A with $(\text{confirm}$, $sid$, $m_{\text{ct}}$, $\theta_{w+1})$, then $\mathcal{S}$ sends $(\text{conf-ok})$ to $\mathcal{F}_{\mathcal{H}}$ in the name of C.

\subsubsection*{Withdraw from channel hub}
To simulate the withdrawal procedure, we let $\mathcal{S}$ simulate the behaviors of honest parties, contract functionality $\mathcal{F}_{\mathcal{C}}$ and channel hub functionality $\mathcal{F}_{\mathcal{N}}$. 
Again, $\mathcal{S}$ needs to simulate the transfer of funding capacity when successfully withdrawn.
The withdrawal procedure starts from $\mathcal{Z}$ sending $(\text{withdraw}$, $sid$, $\beta_{AC}$, $\mathcal{H}_{\not\subset})$ to both A and C or a corrupted party sending a withdrawal request to $\mathcal{F}_{\mathcal{C}}$.
We note that an honest party cannot prevent malicious one from withdrawing a payment channel.
If A sends $(\text{withdraw}$, $sid$, $\beta_{AC}$, $\mathcal{H}_{\not\subset}$, $c$, $\theta_w)$ to $\mathcal{F}_{\mathcal{C}}$ in round 1, 
then $\mathcal{S}$ sends $(\text{withdrawing})$ to C in the name of $\mathcal{F}_{\mathcal{C}}$. 
Otherwise, $\mathcal{S}$ just terminates the simulation. 
Next round if C sends message $(\text{withdraw}$, $sid$, $\beta_{AC}$, $\mathcal{H}_{\not\subset}$, $c$, $\theta_w)$ to $\mathcal{F}_{\mathcal{C}}$, 
then $\mathcal{S}$ sends $(\text{withdraw}$, $sid$, $\beta_{AC}$, $c)$ to $\mathcal{F}_{\mathcal{N}}$ in the name of $\mathcal{F}_{\mathcal{C}}$,
sends $(\text{transfer}$, $sid$, $\alpha_{\mathcal{N}}$, $\alpha_{\mathcal{C}}$, $c)$ to $\mathcal{F}_{\mathcal{L}}$ in the name of $\mathcal{F}_{\mathcal{N}}$, 
and outputs $(\text{withdrawed})$.
If C does not reply with $(\text{withdraw})$ in round 2, then $\mathcal{S}$ postpones the same actions until round 3.

\subsubsection*{Close a payment channel} 
To simulate the closing procedure, we let $\mathcal{S}$ simulate the behaviors of honest parties and the contract functionality $\mathcal{F}_{\mathcal{C}}$.
The closing procedure starts from $\mathcal{Z}$ sending $(\text{close}, sid, \beta_{AC})$ to both A and C or a corrupted party trying to close the payment channel unilaterally.
We note that an honest party still cannot prevent malicious one from closing a payment channel.
If A sends a signed message $(\text{close}, sid, \beta_{AC}, c, \theta_{w_A})$ to $\mathcal{F}_{\mathcal{C}}$ in round 1, 
then $\mathcal{S}$ sends $(\text{closing})$ to C in the name of $\mathcal{F}_{\mathcal{C}}$. 
Otherwise, $\mathcal{S}$ terminates the simulation.
In round 3, $\mathcal{S}$ sends $(\text{transfer}, sid, \alpha_{\mathcal{C}}, A, x_a)$ and $(\text{transfer}, sid, \alpha_{\mathcal{C}}, C, x_c)$ to $\mathcal{F}_{\mathcal{L}}$ in the name of $\mathcal{F}_{\mathcal{C}}$, and outputs $(\text{closed})$.
In case when C sends $(\text{close}, sid, \beta_{AC}, c, \theta_{w_C})$ to $\mathcal{F}_{\mathcal{C}}$ in round 2, then $\mathcal{S}$ does the same actions in round 2.
The value of $x_a$ and $x_c$ depends on the behaviors of both A and C.
As mentioned above, if there exist irrational parties willing to sustain financial loses in order to cause the other to lose its funds, 
we just let $\mathcal{S}$ to move funds from the account of the dishonest party to the honest one using the \textit{transfer} interface.

\end{IEEEproof}

\end{document}